\documentstyle[12pt,epsf]{article}

\catcode`@=11
\def\chkspace{%
  \relax   
  \begingroup\ifhmode\aftergroup\dochksp@ce\fi\endgroup}
\def\dochksp@ce{%
  \unskip              
  \futurelet\chkspct@k\d@chkspc  
}
\def\d@chkspc{%
  \let\nxtsp@ce=\relax
  \ifx\chkspct@k.\else     
    \ifx\chkspct@k,\else
      \ifx\chkspct@k;\else
        \ifx\chkspct@k!\else
          \ifx\chkspct@k?\else
            \ifx\chkspct@k:\else
              \ifx\chkspct@k)\else
              \ifx\chkspct@k(\else
                \ifx\chkspct@k]\else
                  \ifx\chkspct@k-\else
                    \ifx\chkspct@k\egroup\else  
                      \let\nxtsp@ce=\put@space  
                    \fi
                  \fi
                \fi
              \fi
              \fi
            \fi
          \fi
        \fi
      \fi
    \fi
  \fi
  \nxtsp@ce
}
\def\put@space{$\;$}
\catcode`@=12

\def\ra{{$\rightarrow$}\chkspace}

\def\etal{{\it et al.}\chkspace}

\def\adhoc{{\it ad hoc}\chkspace}

\def\eg{{\it eg.}\chkspace}

\def\apriori{{\it a priori}\chkspace}

\def\ep{{e$^+$e$^-$}\chkspace}

\def\gluino{\relax\ifmmode \tilde{g} \else $\tilde{g}$ \fi\chkspace}

\chkspace
\chkspace
\def\m0{$M_{0}$}\chkspace
\def\m0m{$M_{0}MAX$}\chkspace
\chkspace
\chkspace

\def\bbrm{\relax\ifmmode {\rm b}\bar{\rm b}
       \else ${\rm b}\bar{\rm b}$ \fi\chkspace}
\def\bb{$b\bar{b}$ \chkspace}
\def\ccrm{\relax\ifmmode {\rm c}\bar{\rm c}
       \else ${\rm c}\bar{\rm c}$ \fi\chkspace}
\def\cc{$c\bar{c}$ \chkspace}
\def\ss{$s\bar{s}$ \chkspace}
\def\uu{\relax\ifmmode {\rm u}\bar{\rm u}
       \else ${\rm u}\bar{\rm u}$ \fi\chkspace}
\def\dd{\relax\ifmmode {\rm d}\bar{\rm d}
       \else ${\rm d}\bar{\rm d}$ \fi\chkspace}

\def\qqg{\relax\ifmmode {\rm q}\overline{\rm q}{\rm g}
\else q$\overline{\rm q}$g \fi\chkspace}

\def\afb{\relax\ifmmode A_{FB} \else
{{$A_{FB}$}}\fi\chkspace}
\def\afbb{\relax\ifmmode A_{FB}^b \else
{{$A_{FB}^b$}}\fi\chkspace}
\def\pafb{\relax\ifmmode \tilde{A}_{FB} \else
{{$\tilde{A}_{FB}$}}\fi\chkspace}
\def\pafbb{\relax\ifmmode \tilde{A}_{FB}^b \else
{{$\tilde{A}_{FB}^b$}}\fi\chkspace}

\def\pafbzo{\relax\ifmmode \tilde{A}_{FB}|_{O(0)} \else
{{$\tilde{A}_{FB}|_{O(0)}$}}\fi\chkspace}
\def\pafbfo{\relax\ifmmode \tilde{A}_{FB}|_{\oalp} \else
{{$\tilde{A}_{FB}|_{\oalp}$}}\fi\chkspace}
\def\pafbso{\relax\ifmmode \tilde{A}_{FB}|_{\oalpsq} \else
{{$\tilde{A}_{FB}|_{\oalpsq}$}}\fi\chkspace}
\def\pafbto{\relax\ifmmode \tilde{A}_{FB}|_{\oalpc} \else
{{$\tilde{A}_{FB}|_{\oalpc}$}}\fi\chkspace}

\def\pafbbzo{\relax\ifmmode \tilde{A}_{FB}^b|_{O(0)} \else
{{$\tilde{A}_{FB}^b|_{O(0)}$}}\fi\chkspace}
\def\pafbbfo{\relax\ifmmode \tilde{A}_{FB}^b|_{\oalp} \else
{{$\tilde{A}_{FB}^b|_{\oalp}$}}\fi\chkspace}
\def\pafbbso{\relax\ifmmode \tilde{A}_{FB}^b|_{\oalpsq} \else
{{$\tilde{A}_{FB}^b|_{\oalpsq}$}}\fi\chkspace}
\def\pafbbto{\relax\ifmmode \tilde{A}_{FB}^b|_{\oalpc} \else
{{$\tilde{A}_{FB}^b|_{\oalpc}$}}\fi\chkspace}

\def\afbo0{\tilde{A}_{FB}|_{O(0)}}
\def\afbo1{\tilde{A}_{FB}|_{\oalp}}
\def\afbo2{\tilde{A}_{FB}|_{\oalpsq}}
\def\afbo3{\tilde{A}_{FB}|_{\oalpc}}

\def\lam{\relax\ifmmode \Lambda_{\overline{MS}}
       \else {{$\Lambda_{\overline{MS}}$}}\fi\chkspace}
\def\lamuds{\relax\ifmmode \Lambda^{(3)}_{\overline{MS}}
       \else {{$\Lambda^{(3)}_{\overline{MS}}$}}\fi\chkspace}
\def\lamudsc{\relax\ifmmode \Lambda^{(4)}_{\overline{MS}}
       \else $\Lambda^{(4)}_{\overline{MS}}$\fi\chkspace}
\def\lamudscb{\relax\ifmmode \Lambda^{(5)}_{\overline{MS}}
       \else $\Lambda^{(5)}_{\overline{MS}}$\fi\chkspace}

\def\alp{\relax\ifmmode \alpha_s\else $\alpha_s$\fi\chkspace}
\def\alpbar{\relax\ifmmode \bar{\alpha_s}
       \else $\bar{\alpha_s}$\fi\chkspace}
\def\alpmz{\relax\ifmmode \alpha_s(M_Z)\else $\alpha_s(M_Z)$\fi\chkspace}
\def\alpmzsq{\relax\ifmmode \alpha_s(M_Z^2)
       \else $\alpha_s(M_Z^2)$\fi\chkspace}

\def\oalp{\relax\ifmmode O(\alpha_s)\else{{O($\alpha_s$)}}\fi\chkspace}
\def\oalpsq{\relax\ifmmode O(\alpha_s^2)
           \else{{O($\alpha_s^2$)}}\fi\chkspace}
\def\oalpc{\relax\ifmmode O(\alpha_s^3)
           \else{{O($\alpha_s^3$)}}\fi\chkspace}
\def\oalpf{\relax\ifmmode O(\alpha_s^4)
           \else{{O($\alpha_s^4$)}}\fi\chkspace}

\def\rb{\relax\ifmmode R_3^b/R_3^{all}
           \else{{$R_3^b/R_3^{all}$}}\fi\chkspace}
\def\rc{\relax\ifmmode R_3^c/R_3^{all}
           \else{{$R_3^c/R_3^{all}$}}\fi\chkspace}
\def\ruds{\relax\ifmmode R_3^{uds}/R_3^{all}
           \else{{$R_3^{uds}/R_3^{all}$}}\fi\chkspace}
\def\ri{\relax\ifmmode R_3^i/R_3^{all}
           \else{{$R_3^i/R_3^{all}$}}\fi\chkspace}
\def\rj{\relax\ifmmode R_3^j/R_3^{all}
           \else{{$R_3^j/R_3^{all}$}}\fi\chkspace}
\def\alpi{\relax\ifmmode \alpha^i_s/\alpha^{all}_s
           \else{{$\alpha^i_s/\alpha^{all}_s$}}\fi\chkspace}

\def\plb{Phys. Lett.\chkspace}
\def\npb{Nucl. Phys.\chkspace}

\def\prl{Phys. Rev. Lett.\chkspace}
\def\prd{Phys. Rev.\chkspace}
\def\zpc{Z. Phys.\chkspace}

\def\z0{{$Z^0$}\chkspace}
\def\Dst{\relax\ifmmode {\rm D}^* \else {D$^*$}\fi\chkspace}
\def\Dpl{\relax\ifmmode {\rm D}^+ \else {D$^+$}\fi\chkspace}
\def\D0{\relax\ifmmode {\rm D}^0 \else {D$^0$}\fi\chkspace}
\def\Kst{\relax\ifmmode {\rm K}^* \else {K$^*$}\fi\chkspace}
\def\K0{\relax\ifmmode {\rm K}^0_s \else {K$^0_s$}\fi\chkspace}
\def\Kpl{\relax\ifmmode {\rm K}^+ \else {K$^+$}\fi\chkspace}
\def\Kstz{\relax\ifmmode {\rm K}^{*0} \else {K$^{*0}$}\fi\chkspace}

\def\ep{{$e^+e^-$}\chkspace}

\def\z0{$Z^0$}

\def\bb{{$b\bar{b}$}\chkspace}
\def\cc{{$c\bar{c}$}\chkspace}

\def\etal{{\it et al.}\chkspace}
\def\adhoc{{\it ad hoc}\chkspace}

\def\qqg{{$q\bar{q}g$}\chkspace}

\renewcommand{\baselinestretch}{1.0}

 1

\catcode`\@=11 
\makeatletter
\def\@seccntformat#1{\csname the#1\endcsname.\hskip 1em}
\makeatother

\begin{document}

\thispagestyle{empty}
\begin{flushright}
{\renewcommand{\baselinestretch}{.75}
  SLAC--PUB--9087\\
February 2002\\
}
\end{flushright}

\vskip 0.5truecm
 
\begin{center}
{\large\bf
 Measurement of the $b$-Quark Fragmentation \\ Function in $Z^0$ Decays$^*$\\
}
\end{center}
\vspace {0.4cm}

\begin{center}
 {\bf The SLD Collaboration$^{**}$}\\
Stanford Linear Accelerator Center \\
Stanford University, Stanford, CA~94309
\end{center}
 
\vspace{0.5cm}
 
\begin{center}
{\bf ABSTRACT }
\end{center}

\noindent
We present a measurement 
of the $b$-quark inclusive fragmentation function in $Z^{0}$ 
decays using a novel kinematic $B$-hadron energy 
reconstruction technique.  The measurement was performed
using 350,000 hadronic \z0 events recorded in the SLD experiment 
at SLAC between 1997 and 1998.  The small and stable SLC beam spot 
and the CCD-based vertex detector were used to reconstruct 
$B$-decay vertices with high efficiency and purity, 
and to provide precise measurements of the kinematic
quantities used in this technique.  We measured the $B$ energy 
with good efficiency and resolution over the full kinematic range.  
We compared the scaled $B$-hadron energy distribution with 
models of $b$-quark fragmentation and with
several \adhoc functional forms. A number of models and functions 
are excluded by the data.  The average scaled energy of 
weakly-decaying $B$ hadrons was measured to be 
$<x_b>$ = 0.709 $\pm$ 0.003 (stat) $\pm$ 0.003 (syst) $\pm$ 0.002 (model).

\vskip 2truecm

\centerline{\it Submitted to Physical Review D}

\vfill
{\footnotesize
$^*$ Work supported by Department of Energy contract DE-AC03-76SF00515 (SLAC).}

\eject

\rm  
\section{Introduction}
\noindent 
The production of heavy hadrons ($H$) in \ep annihilation provides a
laboratory for the study of heavy-quark ($Q$) jet fragmentation. This is 
commonly characterised in terms of the observable 
$x_{H}$ $\equiv$ $2E_H/\sqrt{s}$, where
$E_H$ is the energy of a $B$ or $D$ hadron containing a $b$ or $c$ quark,
respectively, and $\sqrt{s}$ is the c.m. energy. In contrast to light-quark
jet fragmentation one expects~\cite{Bj} the distribution of $x_{H}$, 
$D(x_{H})$, to peak at an $x_{H}$-value significantly above 0. 
Since the hadronisation process is intrinsically non-perturbative $D(x_{H})$ 
cannot be calculated directly using perturbative Quantum Chromodynamics
(QCD). However, the distribution of the closely-related variable
$x_{Q}$ $\equiv$ 2$E_Q/\sqrt{s}$ can be calculated
perturbatively \cite{collins,mn,bcfy,dkt} and related, via model-dependent
assumptions, to the observable quantity $D(x_{H})$; a number of such
models of heavy-quark fragmentation have been proposed
\cite{kart,bowler,pete,lund}. Measurements of $D(x_{H})$ thus serve to
constrain both perturbative QCD and the model predictions. 
Furthermore, the measurement of $D(x_{H})$ at different c.m. energies
can be used to test QCD evolution, and comparison of $D(x_{B})$
with $D(x_{D})$ can be used to test heavy-quark symmetry~\cite{jaffe}. 
Finally, the uncertainty on the forms of $D(x_{D})$ and $D(x_{B})$
must be taken into account in studies of the production and decay of heavy
quarks, see \eg~\cite{heavy}; more accurate measurements of these forms 
will allow increased precision in tests of the electroweak heavy-quark sector.

We have measured the inclusive weakly-decaying $B$-hadron scaled energy distribution
$D(x_{B})$ in $Z^0$ decays. Earlier studies \cite{early} 
used the momentum spectrum of the lepton from semi-leptonic $B$ decays to 
constrain the mean value $<x_{B}>$ and found it to be approximately
$0.70$; this is in agreement with the results of similar studies at $\sqrt{s}$
= 29 and 35 GeV~\cite{petra}. In more recent
analyses~\cite{aleph95,shape,sldbfrag} 
$D(x_{B})$ has been measured by reconstructing $B$ hadrons via their
$B$ \ra $DlX$ decay mode. In this case the reconstruction efficiency is
intrinsically low due to the small branching ratio for $B$ hadrons to decay into
the high-momentum leptons used in the tag.  Also,
the reconstruction of the $B$-hadron energy using calorimeter information 
usually has poor resolution for low $B$ energy, resulting
in poor sensitivity to the shape of the distribution in this region.

We present the results of a new method for reconstructing
$B$-hadron decays and the $B$ energy inclusively, using only charged tracks,
in the SLD experiment at SLAC.
We used the upgraded charge-coupled device (CCD) vertex detector, 
installed in 1996, to reconstruct $B$-decay vertices with high 
efficiency and purity.  Combined with the micron-size SLC interaction point
(IP), precise vertexing allowed us to reconstruct accurately
the $B$ flight direction and hence 
the transverse momentum of tracks associated with 
the vertex with respect to this direction.  
Using the transverse momentum and 
the total invariant mass of the associated tracks, an upper limit
on the mass of the missing particles was found for each 
reconstructed $B$-decay vertex, and was used to solve for the longitudinal 
momentum of the missing particles, and hence for the energy 
of the $B$ hadron. In order 
to improve the $B$ sample purity and the reconstructed $B$-hadron energy 
resolution, $B$ vertices with low missing mass were selected.
The method is described in Section 3. In Section 4
we compare our reconstructed $D(x_B)$ with the predictions 
of heavy-quark fragmentation models.  We also test several functional 
forms for this distribution.  In Section 5 we describe the unfolding procedure 
used to derive our estimate of the true underlying $D(x_B)$. In Section 6 we discuss the 
systematic errors.  In Section 7 we summarize the results.
Our measurement based on a data sample one-third the size of that used
here is reported in Refs.~\cite{sldbfragprl,dan}.

\section{Apparatus and Hadronic Event Selection}
 
\noindent
This analysis is based on roughly 350,000 hadronic events produced in 
\ep annihilations at a mean center-of-mass energy of $\sqrt{s}=91.28$ GeV
at the SLAC Linear Collider (SLC), and recorded in the SLC Large Detector
(SLD) in 1997 and 1998. 
A general description of the SLD can be found elsewhere~\cite{sld}.
The trigger and initial selection criteria for hadronic $Z^0$ decays are 
described in Ref.~\cite{sldalphas}.
This analysis used charged tracks measured in the Central Drift
Chamber (CDC)~\cite{cdc} and in the upgraded Vertex Detector (VXD3)~\cite{vxd}.
Momentum measurement was provided by a uniform axial magnetic field of 0.6T.
The CDC and VXD3  give a momentum resolution of
$\sigma_{p_{\perp}}/p_{\perp}$ = $0.01 \oplus 0.0026p_{\perp}$,
where $p_{\perp}$ is the track momentum transverse to the beam axis in
GeV/$c$. In the plane normal to the beamline 
the centroid of the micron-sized SLC interaction point (IP) was reconstructed from tracks
in sets of approximately thirty sequential hadronic \z0 decays with a precision 
of $\sigma_{IP}^{r\phi}\simeq4\pm2$ $\mu$m.  The IP position along the 
beam axis was determined event by event using charged tracks with 
a resolution of $\sigma_{IP}^z$ $\simeq$ 20 $\mu$m.
Including the uncertainty on the IP position, the resolution on the 
charged-track impact parameter ($d$) projected in the plane perpendicular
to the beamline was 
 $\sigma_{d}^{r\phi}$ = 8$\oplus$33/$(p\sin^{3/2}\theta)$ $\mu$m,
and the resolution in the plane containing the beam axis was  
 $\sigma_{d}^{z}$ = 10$\oplus$33/$(p\sin^{3/2}\theta)$ $\mu$m,
where
$\theta$ is the track polar angle with respect to the beamline.
The event thrust axis~\cite{thrust} was calculated using energy clusters
measured in the Liquid Argon Calorimeter~\cite{lac}. 

A set of cuts was applied to the data to select well-measured tracks
and events well contained within the detector acceptance.
Charged tracks were required to have a distance of
closest approach transverse to the beam axis within 5 cm,
and within 10 cm along the axis from the measured IP,
as well as $|\cos \theta |< 0.80$, and $p_\perp > 0.15$ GeV/c.
Events were required to have a minimum of seven such tracks,
a thrust-axis  polar angle w.r.t. the beamline, $\theta_T$,
within $|\cos\theta_T|<0.71$, and
a charged visible energy $E_{vis}$ of at least 20~GeV,
which was calculated from the selected tracks, which were assigned the charged pion mass. 
The efficiency for selecting a well-contained $Z^0 \rightarrow q{\bar q}(g)$
event was estimated to be above 96\% independent of quark flavor. The
selected sample comprised 218,953 events, with an estimated
$0.10 \pm 0.05\%$ background contribution dominated
by $Z^0 \rightarrow \tau^+\tau^-$ events.

For the purpose of estimating the efficiency and purity of the $B$-hadron
selection procedure we made use of a detailed Monte Carlo (MC) simulation 
of the detector.
The JETSET 7.4~\cite{jetset} event generator was used, with parameter
values tuned to hadronic \ep annihilation data~\cite{tune},
combined with a simulation of $B$ hadron decays
tuned~\cite{sldsim} to $\Upsilon(4S)$ data and a simulation of the SLD
based on GEANT 3.21~\cite{geant}.
Inclusive distributions of single-particle and event-topology observables
in hadronic events were found to be well described by the
simulation~\cite{sldalphas}. Uncertainties in the simulation 
were taken into account in the systematic errors (Section~\ref{sec:sys}). 

\noindent
\section{$B$-Hadron Selection and Energy Measurement}

\subsection{$B$-Hadron Selection}

The $B$ sample for this analysis was selected using a `topological vertexing'
technique based on the detection and measurement of charged tracks,
which is described in detail in Ref.~\cite{zvnim}. 
Each hadronic event was divided into two hemispheres by a plane perpendicular
to the thrust axis.
In each hemisphere the vertexing algorithm was applied to 
the set of `quality' tracks having
(i) at least 23 hits in the CDC and 2 hits in VXD3; 
(ii) a combined CDC and VXD3 track fit quality of $\chi^{2}/N_{dof}< $8;
(iii) a momentum ($p$) in the range 0.25$<p<$55 GeV/$c$,
(iv) an impact parameter projection in the $r-\phi$ plane of less than 0.3~cm, 
and a projection along the $z$ axis of less than 1.5~cm;
(v) an $r-\phi$ impact parameter error no larger than 250 $\mu$m. 

Vertices consistent with photon conversions or $K^{0}$ and $\Lambda^0$ decays 
were discarded.
In hemispheres containing at least one found vertex the
vertex furthest from the IP was retained 
as the `seed' vertex.  
Those events were retained which contained a seed vertex separated from the IP 
by between
0.1~cm and 2.3~cm. The lower bound reduces contamination from non-$B$-decay
tracks and backgrounds from light-flavor events, and the upper bound
reduces the background from particle interactions with the beam 
pipe. A sample of 76,421 event hemispheres was selected.

In each hemisphere, a 
vertex axis was defined as the straight line joining the IP to
the vertex, which was located at a distance $D$ from the IP.  
For each quality track not directly associated with the vertex,
the distance of closest approach to the vertex axis, $T$,
and the distance from the IP along the vertex
axis to the point of closest approach, $L$, were calculated. 
Tracks satisfying $T<1$~mm and $L/D>0.3$ were added to the vertex.
These $T$ and $L$ cuts were chosen to minimize false track associations 
to the seed vertex, since typically the addition of
a false track has a much greater
kinematic effect than the omission of a genuine $B$-decay track, and hence 
has more effect on the reconstructed $B$-hadron energy.  
Our Monte Carlo studies show that, on average, this procedure 
attaches 0.85 tracks to each seed vertex, 91.9\% of the tracks 
from tagged true $B$ decays are associated
with the resulting vertices, and 98.0\% of the vertex tracks are from true
$B$ decays.  

The large masses of the $B$ hadrons relative to light-flavor hadrons 
make it possible to distinguish $B$-hadron decay vertices from those 
vertices found in events of primary light flavor using the vertex invariant 
mass, $M$. However, due to the effect of those particles missed from being 
associated with the vertex, which are mainly neutrals,
$M$ cannot be fully determined.  
In the {\em rest} frame of the decaying $B$  hadron, $M$ can be written  
\begin{equation}
M=\sqrt{M_{ch}^{2}+P_{t}^{2}+P_{chl}^{2}}+\sqrt{M_{0}^{2}+P_{t}^{2}
\label{eqn:vertexmass}
+P_{0l}^{2}}
\end{equation}
where $M_{ch}$ and $M_{0}$ are the total invariant masses of the set of 
vertex-associated tracks and the set of missed particles, respectively.
$P_{t}$ is the momentum sum, transverse to the $B$ flight 
direction, of the vertex-associated tracks,
which, by momentum conservation, is identical to the transverse momentum sum of 
the missed particles.  $P_{chl}$ and $P_{0l}$ are 
the respective momentum sums along the $B$ flight direction.  
In the $B$ {\em rest} frame, $P_{chl} = P_{0l}$.  
Using the set of vertex-associated charged tracks, we calculated
the total momentum vector ${\vec{P}}_{ch}$ and its component transverse to the
flight direction $P_t$, and the total energy $E_{ch}$ 
and invariant mass $M_{ch}$, assuming the charged-pion mass for each track.
The lower bound for the mass of the decaying hadron, 
the `$P_{t}$-corrected vertex mass', 
\vspace{-0.2cm}
\begin{equation}
M_{Pt} = \sqrt{M_{ch}^{2}+P_{t}^{2}} + |P_{t}|
\label{eqn:masspt}
\end{equation}
was used as the variable for selecting $B$ hadrons.
Our simulations show that the majority of non-$B$ vertices have
 $M_{Pt}$ less than 2.0 GeV/$c^{2}$.  
However, occasionally the measured $P_t$ may fluctuate to a 
much larger 
value than the true $P_t$, causing some charm-decay vertices to have $M_{Pt}$ 
larger than 2.0 GeV/$c^{2}$.  
To reduce this contamination, we calculated the `minimum $P_t$' by 
allowing the IP and the vertex to float to any pair 
of locations within the respective one-sigma error ellipsoids. 
We substituted the minimum $P_t$ in Equation~(\ref{eqn:masspt}) and 
used this modified $M_{Pt}$ as our 
variable for selecting $B$ hadrons~\cite{sldrb98}.  

\begin{figure}[ht]	
\epsfysize5.5 in
\epsfxsize5.5 in
\leavevmode
\epsfbox{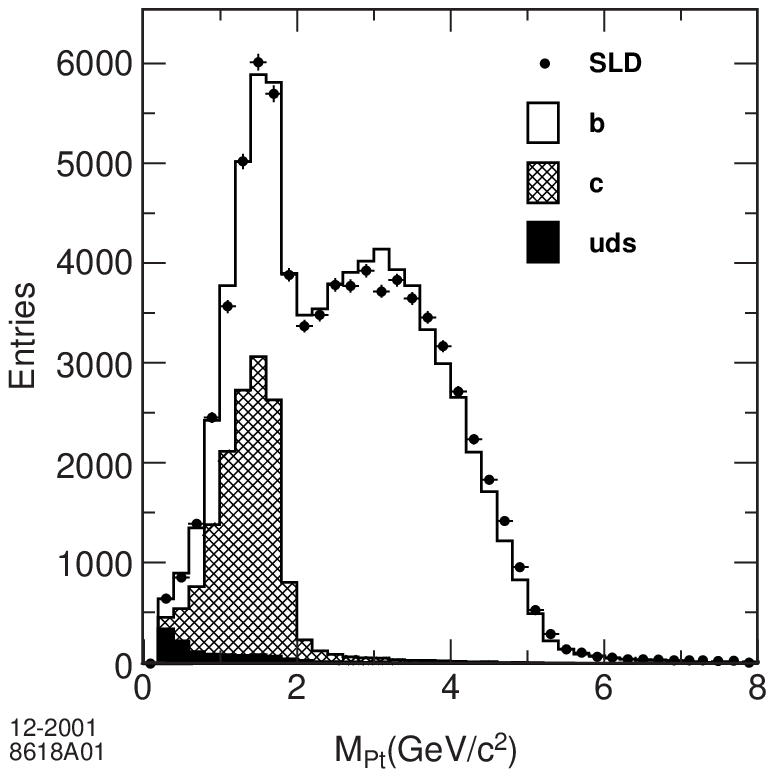}
\vskip -.2 cm
\caption[]{
Distribution of the reconstructed $P_{t}$-corrected vertex mass (points).  
The simulated distribution is also shown (histogram)
in which the flavor composition is indicated:
$b$ (open), $c$ (cross hatched), and $uds$ (dark shaded). 
}
\label{mptm}
\end{figure}

Figure~\ref{mptm} shows the distribution of $M_{Pt}$ 
for the selected sample of hemispheres containing a 
vertex, and the corresponding simulated distribution.  
$B$-hadron candidates were selected by requiring 
$M_{Pt}$ $>$ 2.0 GeV/$c^{2}$.  We further required 
$M_{Pt} \leq 2 \times M_{ch}$ to reduce the contamination from fake
vertices in light-quark events~\cite{sldrb98}.
A total of 42,093 hemispheres were selected, 
with an estimated efficiency for selecting a true $B$-hemisphere 
of 43.7\%, and a sample purity of 98.2\%.  The contributions from 
light-flavors in the sample were 0.34\% for primary $u,d$ and $s$ hemispheres
and 1.47\% for $c$ hemispheres.                                      

\subsection{$B$-Hadron Energy Measurement}

The energy of each $B$ hadron, $E_{B}$, can be expressed as
the sum of the reconstructed-vertex energy, $E_{ch}$, 
and the energy of those true $B$-decay particles that were missed from the vertex, $E_{0}$.  
$E_{0}$ can be written 
\begin{equation}
  E_{0}^{2} =  M_{0}^{2} + P_{t}^{2} + P_{0l}^{2} 
\label{eqn:e0}
\end{equation}
The two unknowns, $M_{0}$ and $P_{0l}$, must be found in order 
to obtain $E_{0}$.
One kinematic constraint can be obtained by imposing the $B$-hadron mass, $M_B$, 
on the vertex.
From Equation~(\ref{eqn:vertexmass}) we 
derive the following inequality,
\begin{equation}
  \sqrt{M_{ch}^2 + P_{t}^2} + \sqrt{M_{0}^2 + P_{t}^2} \leq M_{B}, 
\label{massineq}
\end{equation}
where equality holds in the limit that
both $P_{0l}$ and $P_{chl}$ vanish in the $B$-hadron {\em rest} frame.
Equation~(\ref{massineq}) effectively sets an upper bound on 
$M_{0}$, $M_{0max}$:
\begin{equation}
M_{0max}^{2}=M_{B}^2 - 2M_{B}\sqrt{M_{ch}^2+P_{t}^2} + M_{ch}^2. 
\label{m0maxeqn}
\end{equation}
The lower bound is zero. Hence
\begin{equation}
   0\leq M_{0}^{2}\leq M_{0max}^{2},
\end{equation}
and we expect to obtain
a good estimate of $M_{0}$, and therefore of the $B$-hadron energy,  
when $M_{0max}^{2}$ is small. 

\begin{figure}[ht]	
\epsfysize6.3 in
\epsfxsize6.3 in
\leavevmode
\epsfbox{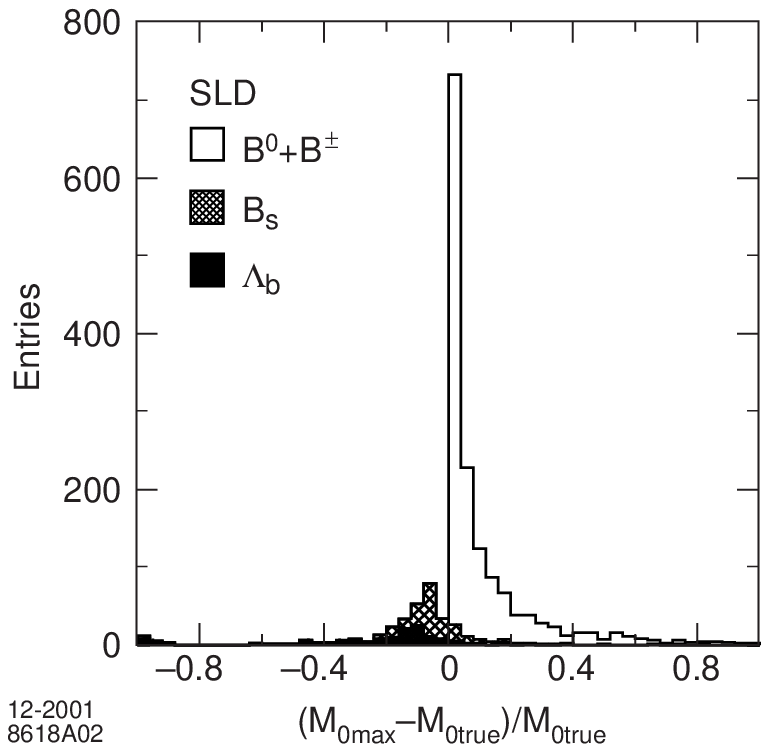}
\vskip -.2 cm
\caption[]{
The relative deviation of the maximum missing mass from the true 
missing mass for simulated $B$ hadron decays;
the contributions from different $B$ species are indicated separately: 
$B^{0}$ and $B^{\pm}$ (open), 
$B_{s}^{0}$ (cross-hatched), and $\Lambda_{b}$ (dark shaded).
}
\label{m0max_m0}
\end{figure}

\begin{figure}[ht]	
\epsfysize5.5 in
\epsfxsize5.5 in
\leavevmode
\epsfbox{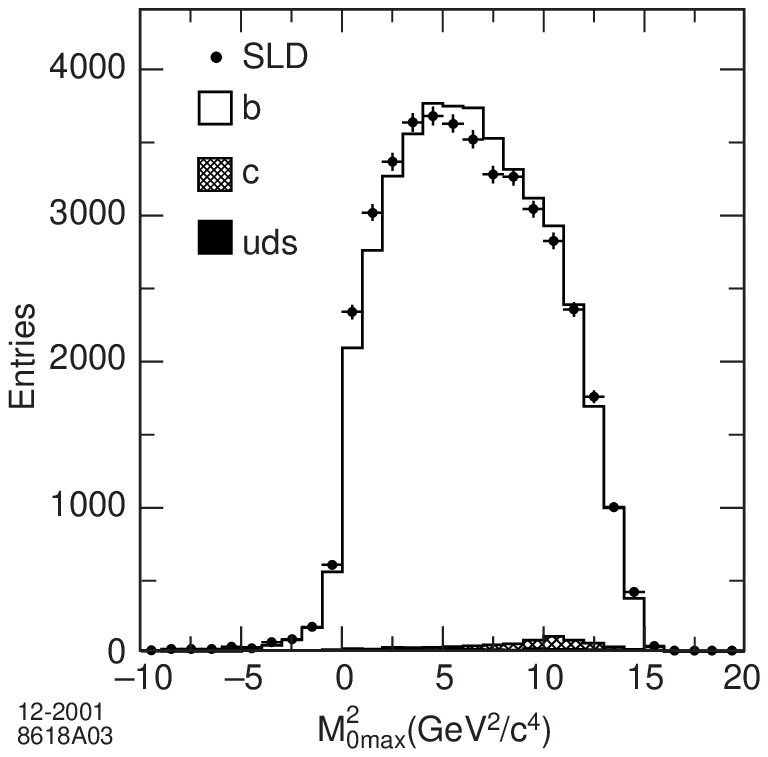}
\vskip -.2 cm
\caption[]{
Distribution of the reconstructed $M_{0max}^{2}$ for the selected 
vertices (points). 
The simulated distribution is also shown (histogram)
in which the flavor composition is indicated:
$b$ (open), $c$ (cross hatched), and $uds$ (dark shaded). 
}
\label{m0max_after}
\end{figure}

We used our simulation to study this issue.
We find that the true value of 
$M_{0}$ tends to cluster near its maximum value $M_{0max}$.
Figure~\ref{m0max_m0} shows 
the relative deviation of $M_{0max}$ from the true $M_0$
for all $B$ hadrons, assuming $M_{B}=$ 5.28 GeV/$c^{2}$ in Eq.~(\ref{m0maxeqn}). 
Although approximately 20\% of the $B$ hadrons are $B^{0}_{s}$ and 
$\Lambda_{b}$, which have larger masses than the $B^0$ and $B^{\pm}$, 
the values of $M_{0max}$ obtained using $M_{B}$=5.28 GeV/$c^{2}$ 
 are typically within about 10\% of $M_0$.
The distribution of the reconstructed $M_{0max}^{2}$ for vertices in
the selected-hemisphere sample is shown in Figure~\ref{m0max_after};
the negative tail is an effect of detector resolution. 
The simulation is in good agreement with the data, and implies that the
non-$B$ background is concentrated at high $M_{0max}^{2}$; this is because
most of the light-flavor vertices have small $M_{Pt}$
and therefore, due to the strong negative correlation between 
$M_{Pt}$ and $M_{0max}$, large $M_{0max}$. 

Because, for true $B$ decays, $M_{0}$ peaks near $M_{0max}$, 
we set $M_{0}^{2}$ = $M_{0max}^{2}$ if $M_{0max}^{2}$ $\geq$0, and
$M_{0}^{2}$ = 0 if $M_{0max}^{2}$ $<$0.
We then calculated $P_{0l}$:
\begin{equation}
   P_{0l} = \frac {\textstyle M_{B}^{2}-(M_{ch}^{2}+P_{t}^{2})-(M_{0}^{2}+P_{t}^{2})}
{\textstyle 2 (M_{ch}^{2}+P_{t}^{2})} P_{chl},
\label{eqn:p0l} 
\end{equation}
and hence $E_{0}$ (Equation~(\ref{eqn:e0})).  
We divided the reconstructed $B$-hadron energy, 
$E_{B}^{rec}=E_{0}+E_{ch}$, by the beam energy, $E_{beam}$, 
to obtain the reconstructed scaled $B$-hadron energy,    
$x_{B}^{rec}$.

The resolution of $x_{B}^{rec}$ depends on both $M_{0max}^{2}$ 
and the true $x_{B}$, $x_{B}^{true}$.  Using our simulation we found that 
vertices that have $M_{0max}^{2}<-1.0 (GeV/c^{2})^{2}$ 
are often poorly reconstructed; we rejected them from further analysis.
Vertices with small values of $|M_{0max}^{2}|$ are typically reconstructed
with better resolution and
an upper cut on $M_{0max}^{2}$ was hence applied.  
For an $x_B$-independent $M_{0max}^{2}$ cut we found that 
the efficiency for selecting $B$ hadrons is roughly linear in $x_{B}$.
In order to obtain an approximately $x_B$-independent selection efficiency
we required:
\begin{equation}
        M_{0max}^{2} < \left\{ 1.1+0.007 (E_{beam}-E_{B}^{rec})+
       4.0 exp[-(E_{B}^{rec}-5.5)/3.5] \right\}^2,
\label{eqn:m0maxcut} 
\end{equation}
where the two \adhoc terms that depend on $E_{B}^{rec}$ 
increase the efficiency at lower $B$-hadron energy.

In addition, in order to reduce 
the light-flavor background, each vertex was required to contain 
at least 3 quality 
tracks with a normalized impact 
parameter greater than 2.  
This cut reduces the dependence of the reconstructed $B$-hadron 
energy distribution on the light-flavor simulation 
in the low-energy region.

\begin{figure}[ht]	
\epsfysize5.5 in
\epsfxsize5.5 in
\leavevmode
\epsfbox{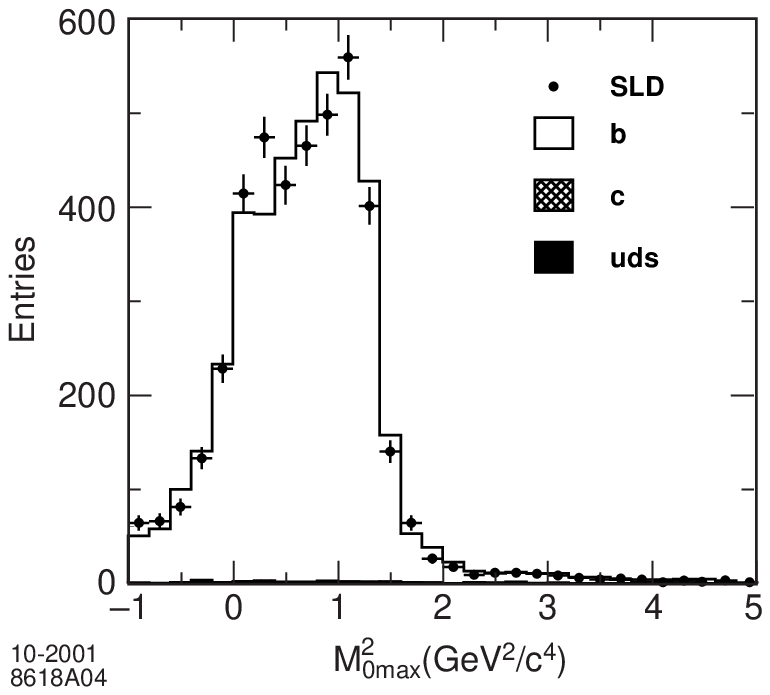}
\vskip -.2 cm
\caption[]{
\label{m0max_before}
Distribution of the reconstructed $M_{0max}^{2}$ for the
final selected sample (see text).
The simulated distribution is also shown (histogram)
in which the flavor composition is indicated:
$b$ (open), $c$ (cross hatched), and $uds$ (dark shaded). 
}
\end{figure}

\begin{figure}[ht]	
\epsfysize6.4 in
\epsfxsize6.4 in
\leavevmode
\epsfbox{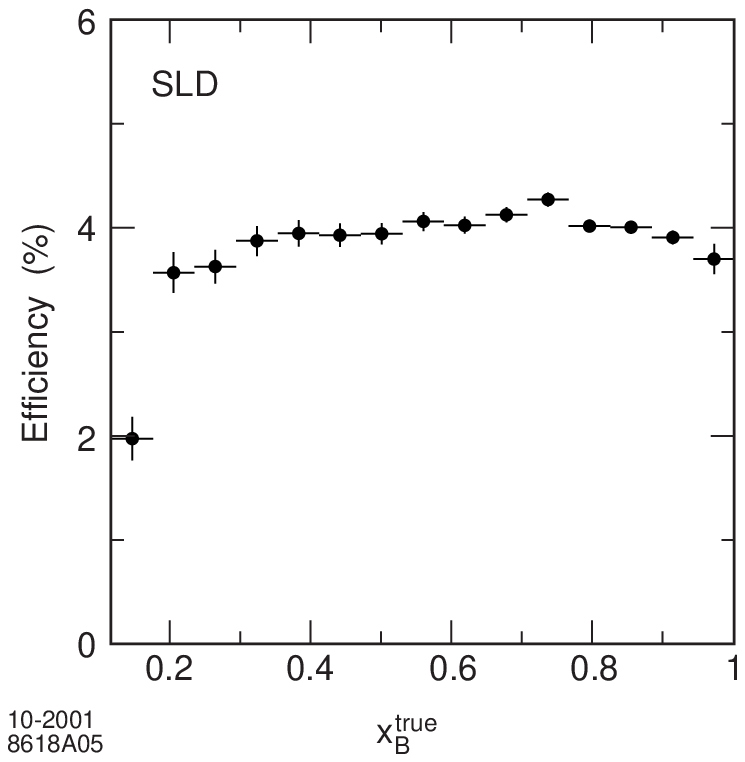}
\vskip -.2 cm
\caption[]{
The simulated efficiency for selecting 
 $B$ hadrons as a function of the 
 true scaled $B$-hadron energy, $x_B^{true}$.
}
\label{efficiency}
\end{figure}

A total of 4,164 hemispheres contained vertices that satisfied 
these selection cuts.  
Figure~\ref{m0max_before} shows the distribution of $M_{0max}^2$; 
the simulation and data are in good agreement.
We calculated that the efficiency for selecting $B$ hadrons is 4.17\% and the 
$B$-hadron purity is 99.0\%, with a $uds$ (charm) background of 0.4\% (0.6\%).
The efficiency as a function of the true $x_{B}$ value, $x_{B}^{true}$, is 
shown in Figure~\ref{efficiency}.  The dependence is weak
except for the lowest $x_B$ region; the efficiency is substantial, 
even just above the kinematic threshold.   

\begin{figure}[ht]	
\epsfysize5.4 in
\epsfxsize5.4 in
\leavevmode
\epsfbox{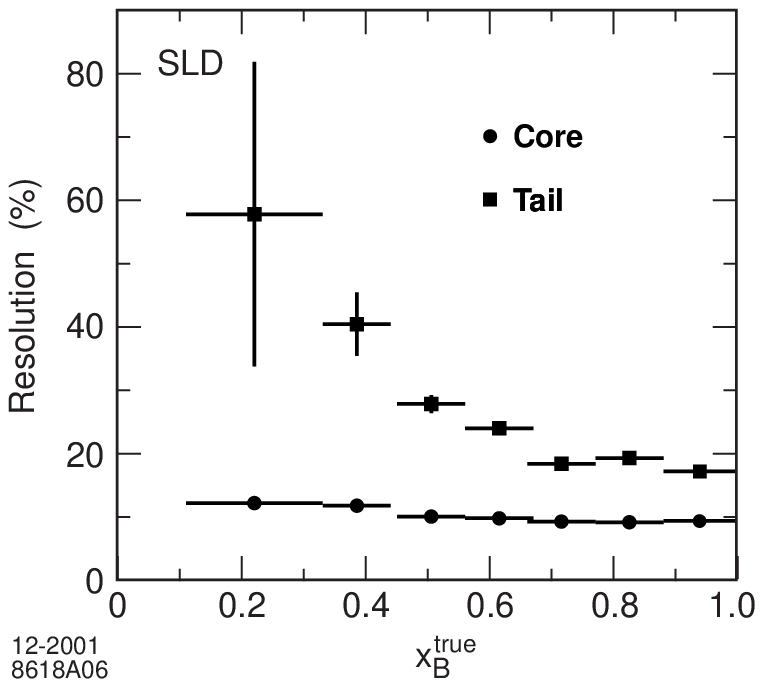}
\vskip -.2 cm
\caption[]{
The fitted core and tail widths (see text) of the $B$-energy resolution as 
a function of the true scaled $B$-hadron energy. 
}
\label{sigmavsx}
\end{figure}

We examined the energy resolution of this technique using simulated events.
The distribution of the normalized difference 
between the true and reconstructed scaled $B$-hadron energies, 
$(x_{B}^{rec}-x_{B}^{true})/x_{B}^{true}$, 
was fitted with the sum of two Gaussians.
A feature of the analysis is that the distribution is symmetric and the fitted
means are consistent with zero.
The fit yields a core width 
  (the width of the narrower Gaussian) of  9.6\% and a tail width 
  (the width of the wider Gaussian) of 21.2\%, with the narrower Gaussian representing a
population fraction of 83.6\%.
Figure~\ref{sigmavsx} shows the core and tail widths as a function 
of $x_{B}^{true}$, where, in order to compare the widths from different $x_B$ bins, 
the ratio between the core and tail populations was fixed to that obtained above.  
The $x_B$-dependence of the resolution is weak. 
The resolution is good even at low $B$ energy, which is an advantage of this energy 
reconstruction technique.

\begin{figure}[ht]	
\epsfysize6.2 in
\epsfxsize6.2 in
\leavevmode
\epsfbox{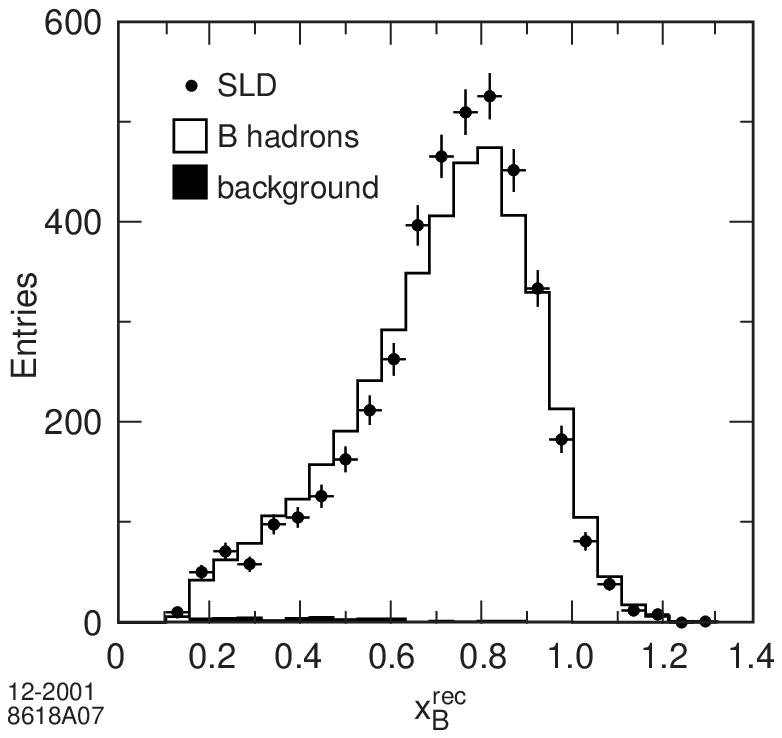}
\vskip -.2 cm
\caption[]{
  Distribution of the reconstructed scaled $B$-hadron energy (points) 
and the default Monte Carlo simulation (histogram).  The solid 
histogram shows the simulated non-$B$ background.
}
\label{xbrec}
\end{figure}

\begin{figure}[ht]	
\epsfysize5.9 in
\epsfxsize5.5 in
\leavevmode
\epsfbox{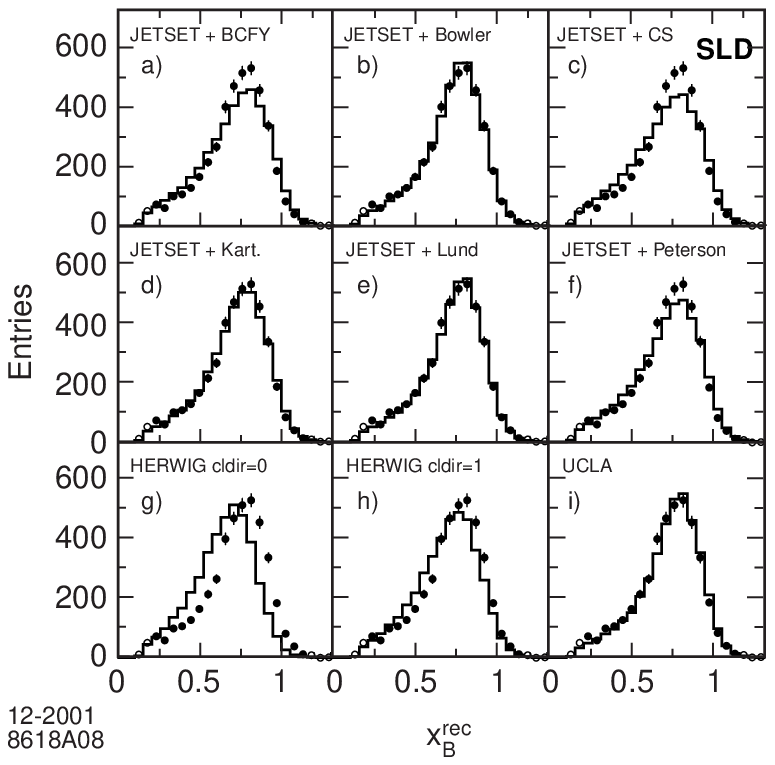}
\vskip -.2 cm
\caption[]{
The background-subtracted distribution of 
reconstructed $B$-hadron energy  (points).
Also shown (histograms) in (a)-(f)  are the predictions of the optimised
models within JETSET (see text). (g), (h) show the predictions of HERWIG,
and (i) of the UCLA model.
Data points excluded from the fit are 
represented by open circles.
}
\label{fig:fragmodel}
\end{figure}

Figure~\ref{xbrec} shows the distribution of the reconstructed scaled 
$B$-hadron energy; the simulated distribution is also shown.
The small non-$B$ background, the high $B$ selection efficiency over 
the full kinematic coverage, and the good energy resolution
combine to give a much improved sensitivity of the data to the underlying 
true {\em shape} of the $B$ energy distribution (see next section).
The distribution of the non-$B$ background
was subtracted bin-by-bin to yield $D^{rec}(x_B^{rec})$, which 
is shown in Fig.~\ref{fig:fragmodel}.

The JETSET event generator used in our simulation is based on a perturbative QCD 
`parton shower' for production of quarks and gluons, together with the 
phenomenological Peterson function~\cite{pete}
(Table~\ref{table:fragmodels})\footnote{We used a value of the 
Peterson function parameter $\epsilon_b$ = 0.006~\cite{sldrb}.} 
to account for the fragmentation of $b$ and $c$ 
quarks into $B$ and $D$ hadrons, respectively,  
within the iterative Lund string hadronisation mechanism~\cite{jetset}.
It is apparent that this simulation does not reproduce the data 
   (Figure~\ref{xbrec}); the $\chi^2$ for the comparison is 70.3 
for 16 bins\footnote{We excluded from the comparison several bins that contained 
very few events; see Section~\ref{subsec:model}.}.

\section{The Shape of the $B$-Hadron Energy Distribution}
\label{sec:shape}

\subsection{Tests of $b$-Quark Fragmentation Models $f(z,\beta)$}
\label{subsec:model}

We tested models of $b$-quark fragmentation.
Since the resulting fragmentation functions are 
usually functions of an experimentally-inaccessible variable $z$, 
\eg $z=(E+p_{\|})_{H}/(E+p_{\|})_Q$ or $z = {p_{\|}}_H / {p_{\|}}_Q$, where $p_{\|}$
represents the hadron momentum along the primary heavy-quark momentum vector, 
it is necessary to use a Monte Carlo generator 
to produce events according to a given input fragmentation function 
$f(z,\beta)$, where $\beta$ represents the set of model arbitrary parameters.

We considered the phenomenological models of 
the Lund group~\cite{lund}, Bowler~\cite{bowler},
Peterson \etal~\cite{pete} and Kartvelishvili \etal~\cite{kart}.  
We also considered the perturbative QCD calculations of Braaten 
\etal~(BCFY)~\cite{bcfy}, and of Collins and Spiller (CS)~\cite{collins}. 
Table~\ref{table:fragmodels} contains a list of the models.
We implemented in turn each fragmentation model in JETSET
and generated events without detector simulation.  
In addition, we tested the UCLA fragmentation model~\cite{ucla} with
default parameter settings, as there is no explicit parameter for controlling the 
$B$-hadron energy. We also tested
the HERWIG~\cite{herwig} event generator, and used both possible settings 
of the parameter switch {\tt cldir}. {\tt cldir}=1 forces the heavy hadron to
continue in the heavy-quark direction in the hadronisation-cluster decay rest frame,
and thereby hardens the fragmentation function. {\tt cldir}=0 suppresses this
feature and yields a softer fragmentation function.

\begin{table}[htb]
\begin{center}
\begin{tabular}{|l|c|c|}
\hline
Model  &  $f(z,\beta)$    &   Reference \\
\hline
BCFY &  $\frac{\textstyle z(1-z)^{2}}{\textstyle [1-(1-r)z]^{6}}[3+{\sum_{i=1}^{4} 
(-z)^{i}f_{i}(r)}]$  & \cite{bcfy} \\
 & & \\
Bowler  & $\frac{\textstyle 1}
{\textstyle z^{(1+r_{b}bm_{\perp}^{2})}}(1-z)^{a}$exp$(-bm_{\perp}^{2}/z)$
      & \cite{bowler} \\
 & & \\
CS &$ (\frac{\textstyle 1-z}{\textstyle z}+\frac{\textstyle (2-z)\epsilon_{b}}{\textstyle 1-z})
(1+z^{2})(1-\frac{\textstyle 1}{\textstyle z}-\frac{\textstyle 
\epsilon_{b}}{\textstyle 1-z})^{-2}$ & \cite{collins} \\
 & & \\
Kartvelishvili  & $z^{\alpha_{b}}(1-z)$ & \cite{kart} \\
 & & \\
Lund  & $\frac{\textstyle 1}{\textstyle z}(1-z)^{a}$exp$(-bm_{\perp}^{2}/z)$
      & \cite{lund} \\
 & & \\
Peterson & $\frac{\textstyle 1}{\textstyle z}(1-\frac{\textstyle 1}{\textstyle z}
-\frac{\textstyle \epsilon_{b}}{\textstyle 1-\textstyle z})^{-2}$   & \cite{pete} \\
\hline
\end{tabular}
\caption{
\label{table:fragmodels} 
$b$-quark fragmentation models used in comparison with the data.  
For the BCFY model, $f_{1}(r)~=~3(3-4r)$, 
$f_{2}(r)~=~12-23r+26r^{2}$, $f_{3}(r)~=~(1-r)(9-11r+12r^{2})$, and 
$f_{4}(r)~=~3(1-r)^{2}(1-r+r^{2})$. 
}
\end{center}
\end{table}

In order to make a consistent comparison of each model 
with the data we adopted the following procedure.  For each model 
starting values of the arbitrary parameters, $\beta$, were assigned 
and the corresponding fragmentation function $f(z,\beta)$ was used to produce the 
scaled weakly-decaying $B$-hadron energy distribution, $D^{true}_{model}(x_{B}^{true},\beta)$ 
{\it before} simulation of the detector.  
The corresponding reconstructed distribution, $D^{rec}_{model}(x_{B}^{rec},\beta)$,  
was derived from the reconstructed distribution generated with our default
model, $D^{rec}_{default}(x_{B}^{rec})$ (Fig.~\ref{xbrec}), 
by weighting events at the generator level with the weight factor 
$D^{true}_{model}(x_{B}^{true},\beta)$/$D^{true}_{default}(x_{B}^{true})$. 
The resulting reconstructed distribution was then compared with the 
data distribution, and the $\chi^2$ value, defined as
\begin{equation}
\chi^2 = \sum_{i=1}^{N} \left( \frac{\textstyle N_{i}^{data} - r N_{i}^{MC} }
{\textstyle \sigma_{i}} \right)^{2}
\label{eqn:chisq}
\end{equation}
was calculated, where $N$ is the number of bins used in the comparison, 
$N_{i}^{data}$ is the number of entries
in bin $i$ in the data distribution, and $N_{i}^{MC}$ is the number of entries 
in bin $i$ in the simulated distribution\footnote{$r$ is the factor by which 
the total number of entries 
in the simulated distribution was scaled to the number of entries in 
the data distribution; $r$ $\simeq$ 1/12.}.
$\sigma_{i}$ is the statistical error on the deviation of the 
observed number of entries for the data from the expected number of 
entries in bin $i$, which can be expressed as
\begin{equation}
\sigma_i^2 = \left( \sqrt{rN_i^{MC}} \right)^2 + 
\left( r\sqrt{N_i^{MC}} \right)^2,
\label{eqn:error}
\end{equation}
where $ \left( \sqrt{rN_{i}^{MC}} \right)^2$ is the expected statistical 
variance on the observed number of entries in bin $i$, 
assuming the model being tested is correct, and 
$ \left( r\sqrt{N_{i}^{MC}} \right)^2 $ is the statistical variance on 
the expected number of entries in bin $i$.  Since the $\chi^{2}$-test is 
not statistically effective for bins with a very small number of 
entries, the third, the fourth, and the last three bins in Figure~\ref{xbrec} 
were excluded from the comparison.

For each model we varied the values of the parameters $\beta$ and repeated the
above procedure.  The minimum $\chi^2$ was found by scanning through 
the input parameter space, yielding
a set of parameters which give an optimal description of the reconstructed
data by the fragmentation model in question.  
The resulting distributions are shown in Figure~\ref{fig:fragmodel}.
Table~\ref{table:modelresult} lists the results of the comparisons.

\begin{table}[htb]
\begin{center}
\begin{tabular}{|l|c|c|c|}
\hline
Model &  $\chi^{2}/dof$    &   Parameters &  $\langle x_{B} \rangle$\\
\hline
JETSET + BCFY
     &  105/16   & $r=0.085$ & 0.694 \\
JETSET + Bowler* &  17/15   & $a=1.4, b=1.2, (r_b=1)$          & 0.709 \\
JETSET + CS &  142/16   & $\epsilon_b=0.003$ & 0.691 \\
JETSET + Kartvelishvili* {\em et al.}  
      &  32/16   & $\alpha_b = 10.0$  &  0.708  \\
JETSET + Lund*   &  17/15   & $a=1.4, b=0.4$          & 0.712 \\
JETSET + Peterson {\em et al.}  & 70/16 & $\epsilon_{b}=0.0055$  &  0.700 \\
HERWIG {\tt cldir}=0  & 1015/17 &   $-$   &  0.632 \\
HERWIG {\tt cldir}=1  &  149/17 &   $-$    &  0.676 \\
UCLA* &  27/17 &   $-$    &  0.718 \\
\hline
\end{tabular}
\caption{
\label{table:modelresult} 
Results of fragmentation model tests.
The minimum $\chi^{2}$, 
number of degrees of freedom, corresponding parameter
values, and the mean value of the corresponding $B$-energy distribution 
are listed. A * indicates those models used below to unfold the data.
}
\end{center}
\end{table}

We conclude that with our resolution and our current data sample, we
are able to distinguish among these fragmentation models.
Within the context of the JETSET fragmentation scheme, the Lund and Bowler 
models are consistent with the data with $\chi^2$ probabilities of
   31\% and 35\%, respectively, the Kartvelishvili model is 
consistent with the data at the 1\% level, while the Peterson, 
BCFY and CS models are 
found to be inconsistent with the data.
The UCLA model is consistent with the data at a level of 6\% 
$\chi^2$ probability.  The HERWIG model with {\tt cldir}=0 is confirmed to 
be much too soft;  using {\tt cldir}=1 results in a harder distribution and a 
substantial improvement, but it is still too soft relative to the data. 

\subsection{Tests of Functional Forms $f(x_B,\lambda)$}
\label{subsec:form}

We considered the more general question of what functional forms, 
$f(x_B,\lambda)$, can be used as estimates of the true 
scaled $B$-energy distribution.
We considered the functional forms of 
the BCFY, CS, Kartvelishvili, Lund, and Peterson 
groups in terms of the variable $x_B^{true}$.
In addition we considered \adhoc generalisations of the Peterson function (`F'),
an 8th-order polynomial (`P8') and a `power' function.  These functions are 
listed in Table~\ref{table:functionalform}. 
Each function vanishes at $x_{B}=0$ and $x_{B}=1$.

\begin{table}[htb]
\begin{center}
\begin{tabular}{|l|c|c|} 
\hline
Function  &  $f(x_B,\lambda)$    &   Reference \\ 
\hline
F & $\frac{\textstyle (1+b(1-x_B))}{\textstyle x_B}(1-\frac{\textstyle c}{\textstyle x_B}
-\frac{\textstyle d}{\textstyle 1-x_B})^{-2}$ & \cite{aleph95} \\
 & & \\
P8 & $x_B(1-x_B)(x_B-x_B^0)(1+{\sum_{i=1}^{5} p_{i}x_B^{i}})$  &   
(see text)  \\
 & & \\
Power  & $x_B^{\alpha}(1-x_B)^{\beta}$  &  (see text) \\
\hline
\end{tabular}
\end{center}
\caption{
Additional \adhoc $B$ energy functional forms used in comparison with the data;
$x_B^0$ = $M_B/E_{beam}$.  
}
\label{table:functionalform}
\end{table}

\begin{figure}[ht]	
\epsfysize5.9 in
\epsfxsize5.5 in
\leavevmode
\epsfbox{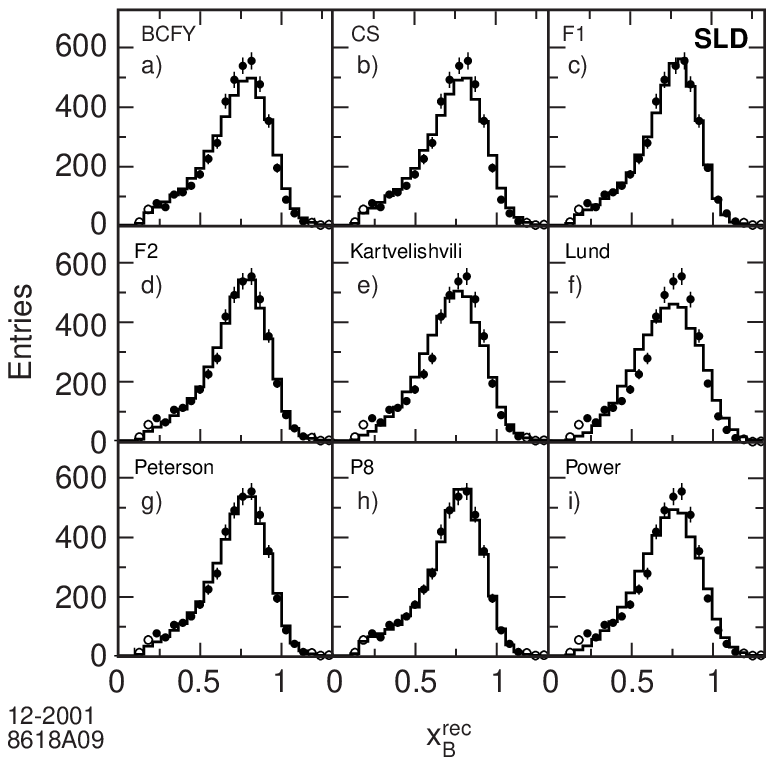}
\vspace{0.1cm}
\vskip -.2 cm
\caption[]{
The background-subtracted distribution of 
reconstructed $B$-hadron energy  (points).
Also shown (histograms) are the predictions of the optimised
functional forms. Data points excluded from the fit are 
represented by open circles.
}
\label{fig:form}
\end{figure}

For each functional form, a testing procedure similar to that described in 
subsection~\ref{subsec:model} was applied. The fitted parameters $\lambda$ 
and the minimum $\chi^2$ values are listed 
in Table~\ref{table:formresult}. The corresponding $D^{rec}_{model}(x_{B}^{rec})$
are compared with the data in Figure~\ref{fig:form}.

\begin{table}[htb]
\begin{center}
\begin{tabular}{|l|c|c|c|}
\hline
Function &  $\chi^{2}/dof$    &   Parameters &  $\langle x_{B} \rangle$\\
\hline
BCFY  &  73/16 & $r=0.248\pm0.007$ & 0.704$\pm$0.003 \\
 & & & \\
CS  
     &  75/16   & $\epsilon_{b}=0.0519\pm0.0036$ & 0.706$\pm$0.003 \\
 & & & \\
Kartvelishvili {\em et al.}  
      &  138/16   & $\alpha_{b}=3.904\pm0.072$  &  0.710$\pm$0.003   \\
 & & & \\
Lund  &  252/15  & $a=1.88\pm0.08$ & 0.715$\pm$0.003 \\
      &            & $bm_{\perp}^{2}=0.32\pm0.05$ &     \\
 & & & \\
Peterson {\em et al.}*  & 31/16 & 
$\epsilon_{b}=0.0382\pm0.0016$ & 0.709$\pm$0.003 \\
 & & & \\
F1* & 20/15  & $c=0.884\pm0.014$ &    0.707$\pm$0.003 \\     
      &            & $d=0.0181\pm0.0015$           &     \\
 & & & \\
F2* & 31/15  & $c=0.976\pm0.029$ & 0.710$\pm$0.003 \\ 
      &            & $d=0.039\pm0.002$           &     \\
 & & & \\
P8*
      & 12/12    &  $p_{1}=-9.99\pm0.25$             &  0.709$\pm$0.003   \\
      &            &  $p_{2}=40.84\pm0.25$    &   \\
            &          & $p_{3}=-82.26\pm0.68$  &   \\
            &          & $p_{4}=80.90\pm0.76$   &   \\
            &          & $p_{5}=-30.60\pm0.54$   &   \\
 & & & \\
Power 
      &  133/15   & $\alpha=3.73\pm0.17$   & 0.713$\pm$0.003  \\
      &            & $\beta=0.84\pm0.07$  &            \\
\hline
\end{tabular}
\caption{
\label{table:formresult} 
Results of the $\chi^{2}$ fit of \adhoc functions to the reconstructed
scaled $B$-hadron energy distribution.  The minimum 
$\chi^{2}$ value, the number of degrees of freedom, the corresponding 
parameter values, and the mean value of the corresponding $B$-energy 
distribution are listed.  Errors are statistical only. 
A * indicates those functions used below to correct the data.
}
\end{center}
\end{table}
\clearpage

Two sets of optimised parameters were found for the generalised
Peterson function F:  
`F1', obtained by setting the parameter $b$ (Table~\ref{table:functionalform}) to infinity, 
behaves like $x_B$ as $x_B$ \ra 0 and $(1-x_B)^3$ as $x_B$ \ra 1 and yields 
a $\chi^2$ probability of 18\%; 
`F2', obtained by setting $b$ to zero, has a $\chi^2$ probability of 1.0\%.  
A constrained polynomial of at least 8th-order was needed to obtain
a $\chi^{2}$ probability greater than 0.1\%.
The Peterson function reproduces the data with a 
$\chi^2$ probability of about 1\%.
The remaining functional forms are found to be inconsistent with the data.
The widths of the BCFY and CS functions are too large to 
describe the data; the Kartvelishvili, Lund and `power'
functions vanish too fast as $x_B$ \ra 0.  
We conclude that, within our resolution and with our 
current data sample, we are able to distinguish among these \adhoc
functional forms.  

\section{Correction of the $B$-Energy Distribution}
\label{sec:correct}

In order to compare our results with those from other experiments and
potential future theoretical predictions it is
necessary to correct $D^{rec}(x_{B}^{rec})$ for the 
effects of detector acceptance, event selection and
analysis bias, as well as for bin-to-bin
migrations caused by the finite resolution of the detector and the
analysis technique. 
Due to the known rapid variation of the \apriori unknown true $B$-energy
distribution at large $x_B$, {\em any} correction procedure will 
necessarily be model-dependent.  
We chose a method that allows explicit 
evaluation of this model-dependence and which gives a very good estimate of the 
true energy distribution using all of the above models or functional 
forms that are consistent with the data.

We applied a $25\times25$ matrix unfolding procedure 
to $D^{rec}(x_{B}^{rec})$ to obtain an estimate of the true distribution 
$D^{true}(x_{B}^{true})$, where $x_{B}^{true}$ refers to the weakly-decaying
$B$ hadron:
\vspace{-0.12cm}
\begin{eqnarray}
D^{true}(x_{B}^{true})\quad=\quad \epsilon^{-1}(x_{B}^{true}) \cdot 
E(x_{B}^{true},x_{B}^{rec}) \cdot D^{rec}(x_{B}^{rec})
\label{eqn:unfold}
\vspace{-0.5cm}
\end{eqnarray}
where $E$ is a
matrix to correct for bin-to-bin migrations, and $\epsilon$ is
a vector representing the efficiency for selecting true $B$-hadron
decays. 
$E$ and $\epsilon$ were calculated from our MC
simulation; the matrix $E$ incorporates a
convolution of the input fragmentation function with the resolution of the
detector.  $E(i,j)$ is the number of vertices with $x_{B}^{true}$ in bin $i$ 
and $x_{B}^{rec}$ in bin $j$, normalized by the total number of vertices 
with $x_{B}^{rec}$ in bin $j$.  

\begin{figure}[ht]	
\epsfysize6.0 in
\epsfxsize5.5 in
\leavevmode
\epsfbox{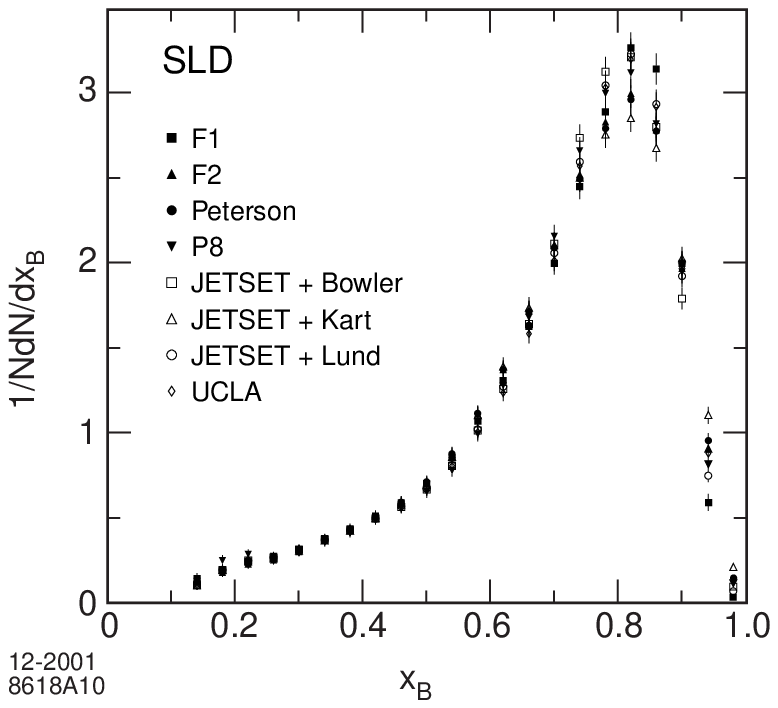}
\vskip -.2 cm
\caption[]{
Distributions of the scaled weakly-decaying $B$-hadron energy unfolded
using different input models or functions (see text).
}
\label{overlay}
\end{figure}

We evaluated $E$ by using in turn the Monte Carlo simulation weighted 
according to each input generator-level {\em true} $B$ energy 
distribution found to be consistent with the data in 
Section~\ref{sec:shape}.  
We considered in turn each of the eight consistent distributions, 
using the optimised parameters listed in Tables~\ref{table:modelresult}
and~\ref{table:formresult}.
The matrix $E$ was then evaluated by examining 
the population migrations of true $B$ hadrons between bins 
of the input scaled $B$ energy, $x_{B}^{true}$, and 
the reconstructed scaled $B$ energy, $x_{B}^{rec}$. 
Using each $D^{true}_{model}(x_{B}^{true})$, the data distribution 
$D^{rec}(x_{B}^{rec})$ was then unfolded
according to Equation~(\ref{eqn:unfold}) to yield $D^{true}(x_{B}^{true})$, 
which is shown for each input fragmentation function in Figure~\ref{overlay}.
It can be seen that the shapes of $D^{true}(x_{B}^{true})$ differ 
systematically among the input scaled $B$-energy distributions.
These differences were used to assign systematic errors.
  
\section{Systematic Errors}
\label{sec:sys}

We considered sources of systematic uncertainty that potentially affect 
our measurement. 
These may be divided into uncertainties in 
modelling the detector and uncertainties on 
experimental measurements serving as
input parameters to the underlying physics modelling. 
For each source of systematic error, the Monte Carlo distribution 
$D^{true}_{default}(x_B^{true})$ was re-weighted and then the resulting 
new reconstructed distribution, $D^{rec}_{new}(x_B^{rec})$, was 
compared with the data $D^{rec}(x_B^{rec})$ 
by repeating the fitting and unfolding procedures described in Sections 4
and 5.  
The differences in both the shape and the mean value of the $x_B^{true}$ 
distribution relative to the default procedure were considered.  

{\it Ad hoc} corrections were applied to the simulations of
four track-related quantities to account for discrepancies w.r.t. the data, namely
the tracking efficiency and the distributions of track $p_{\perp}$,
polar angle and the projection of the impact parameter along the $z$ axis.
In each case a systematic error was assigned (see Table~\ref{table:syst}) using
half the difference between the results obtained with the default and
corrected simulations.

A large number of measured quantities relating to the production and decay
of charm and bottom hadrons are used as input to our simulation. 
In \bb events we considered the uncertainties on: 
the branching fraction for \z0 \ra \bb;
the rates of production of $B^{\pm}$, $B^0$ and $B^0_s$ mesons, 
and $b$ baryons;
the lifetimes of $B$ mesons and baryons;
and the average $B$ hadron decay charged multiplicity.
In \cc events we considered the uncertainties on: 
the branching fraction for \z0 \ra \cc,
the charmed hadron lifetimes,
the charged multiplicity of charmed hadron decays,
the production of  $K^0$ from charmed hadron decays,
and the fraction of charmed hadron decays containing no $\pi^0$s.
We also considered the rate of production of \ss in the jet fragmentation
process, and the production of secondary \bb and \cc from gluon splitting.
The world-average values and their respective uncertainties~\cite{heavy,sldrb} 
were used in our simulation and are listed in
Table~\ref{table:syst}.  Most of these variations affect
the normalisation, but have very little effect on the shape or the mean value.  In no
case do we find a variation that changes our conclusion about which models and
functions are consistent with the data.  The systematic errors on the mean value
are listed in Table~\ref{table:syst}.

\begin{table}[htb]
\begin{center}
\begin{tabular}{|l|c|r|}
\hline
Source                          & Variation & $\delta$ $\langle x_B \rangle$ \\
\hline
 Tracking efficiency correction    & -1.5$\pm$0.75\%            & 0.0007 \\ 
 Impact parameter smearing in $z$  &  9.0$\pm$4.5$\mu$m       & 0.0006 \\
 Track polar angle smearing        &  1.0$\pm$0.5 mrad        & 0.0002 \\ 
 Track $1/p_\perp$ smearing 	   &  0.8$\pm$0.4 MeV$^{-1}$  & 0.0013 \\
\hline
    {\bf Detector total}           &                          & {\bf 0.0016} \\
\hline
 $B^+$ production fraction         &  0.39$\pm$0.11           & $\mp$0.0001 \\
 $B^0$ production fraction         &  0.39$\pm$0.11           & $<$0.0001    \\
 $B_s$ production fraction         &  0.098$\pm$0.0012        & $\pm$0.0003  \\
 $\Lambda_b$ production fraction   &  0.103$\pm$0.018         & $<$0.0001    \\
 $B\!\rightarrow$ charm
    multiplicity and species       &     \cite{dan}           & $\pm$0.0006       \\
 $B\!\rightarrow K^0$ multiplicity &  0.658$\pm$0.066         & $\pm$0.0009       \\
 $B\!\rightarrow \Lambda^0$
       multiplicity                &  0.124$\pm$0.008         & $\pm$0.0002       \\
 $B$ decay $<n_{ch}>$              &   4.955$\pm$0.062        & $^{-0.0004}_{+0.0001}$ \\ 

 $D \rightarrow K^0$ multiplicity  &  \cite{dan}              & $\pm0.0014$ \\
 $D \rightarrow$ no $\pi^0$
                        fraction   &  \cite{dan}              & $\pm$0.0006  \\
 $D$ decay $<n_{ch}>$              &  \cite{dan}              & $\pm$0.0003  \\

 $g \rightarrow b\bar{b}$          & 0.00254$\pm$0.00050 /evt & $\pm$0.0001       \\
 $g \rightarrow c\bar{c}$          & 0.0299$\pm$0.0039  /evt  & $\pm$0.0003       \\

$B^0$ mass                         & 5.2794$\pm$ 0.0005 GeV/$c^2$  & $<$0.0001    \\
$b$, $c$ hadron lifetimes, $R_b$, $R_c$    & \cite{summer}        & $\pm$0.0002       \\
\hline
{\bf Physics total}                &                          & {\bf 0.0020} \\
\hline
{\bf Monte Carlo statistics}       &                          & {\bf 0.0008} \\
\hline
{\bf Total systematic}            &                           & {\bf 0.0027} \\
\hline
\end{tabular}
\caption{
\label{table:syst} 
Uncertainty source, range of variation and size of the resulting systematic error on $<x_B>$.
}
\end{center}
\end{table}

\clearpage

Other relevant systematic effects such as variation of 
the event selection cuts and the assumed $B$-hadron mass were also 
found to be very small.  As a cross-check, we varied the $M_{0max}$ 
cut (Equation~(\ref{eqn:m0maxcut})) used to select the final $B$ sample 
and repeated the
analysis procedure.  In each case, conclusions about the shape of the $B$ 
energy distribution hold.  In each bin, all sources of systematic 
uncertainty were added in quadrature to obtain the total systematic error.

\begin{figure}[ht]	
\epsfysize6.0 in
\epsfxsize5.5 in
\leavevmode
\epsfbox{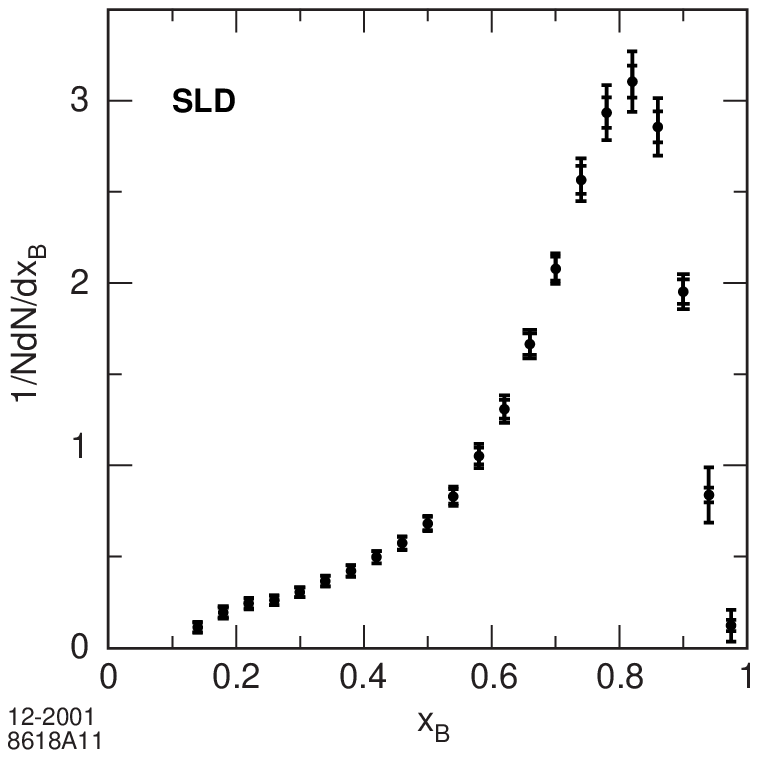}
\vskip -.2 cm
\caption[]{
Final distribution of the weakly-decaying scaled $B$-hadron energy. In each bin the
central value is the average of the eight distributions shown in Fig.~\ref{overlay},
the inner error bar represents the experimental error, and the outer error bar
represents the sum in quadrature of the experimental and unfolding errors.  
}
\label{average}
\end{figure}

The model-dependence of the unfolding procedure was estimated by considering
the envelope of the unfolded results shown in Figure~\ref{overlay}.
Since eight models or functions are consistent with 
the data, in each bin of $x_{B}^{true}$ we calculated the 
average value of these eight unfolded results as well as the r.m.s. 
deviation;  the average was taken as our central
value and the deviation was assigned as the unfolding uncertainty.
Figure~\ref{average} shows the final corrected $x_{B}$ 
distribution $D(x_{B})$. The data are listed in Table~\ref{table:final}
Since two of the eight functions (the Kartvelishvili model and the 
Peterson functional form) are only in marginal agreement with the data,
and the 8th-order polynomial has an unphysical behavior
near $x_B=1$, this r.m.s. may be considered to be a rather reasonable 
envelope within which the true $x_B$ 
distribution is most likely to vary.  The model dependence of this
analysis is significantly smaller than that of previous direct 
$B$-energy measurements, indicating
the enhanced sensitivity of our data to the underlying true energy 
distribution.    

\begin{table}[htb]
\begin{center}
\begin{tabular}{|c|c|c|c|c|c|}
\hline
$x_B$ range       & $D(x_B)$ &  stat. & systematic       & unfolding &    total   \\
\hline
 $  0.00<x_B<0.04$ &  0.000 &  0.000   &  0.000           &   0.000   &    0.000   \\
 $  0.04<x_B<0.08$ &  0.000 &  0.000   &  0.000           &   0.000   &    0.000   \\
 $  0.08<x_B<0.12$ &  0.000 &  0.000   &  0.000           &   0.000   &    0.000   \\
 $  0.12<x_B<0.16$ &  0.110 &  0.029   &  0.004           &   0.014   &    0.034   \\
 $  0.16<x_B<0.20$ &  0.188 &  0.035   &  0.005           &   0.025   &    0.043   \\
 $  0.20<x_B<0.24$ &  0.204 &  0.032   &  0.006           &   0.013   &    0.036   \\
 $  0.24<x_B<0.28$ &  0.213 &  0.027   &  0.008           &   0.010   &    0.030   \\
 $  0.28<x_B<0.32$ &  0.268 &  0.031   &  0.009           &   0.015   &    0.036   \\
 $  0.32<x_B<0.36$ &  0.340 &  0.036   &  0.011           &   0.011   &    0.039   \\
 $  0.36<x_B<0.40$ &  0.398 &  0.037   &  0.012           &   0.010   &    0.041   \\
 $  0.40<x_B<0.44$ &  0.505 &  0.041   &  0.014           &   0.016   &    0.045   \\
 $  0.44<x_B<0.48$ &  0.587 &  0.042   &  0.015           &   0.015   &    0.048   \\
 $  0.48<x_B<0.52$ &  0.677 &  0.044   &  0.016           &   0.011   &    0.050   \\
 $  0.52<x_B<0.56$ &  0.796 &  0.047   &  0.017           &   0.030   &    0.059   \\
 $  0.56<x_B<0.60$ &  0.991 &  0.052   &  0.018           &   0.056   &    0.079   \\
 $  0.60<x_B<0.64$ &  1.241 &  0.058   &  0.018           &   0.070   &    0.092   \\
 $  0.64<x_B<0.68$ &  1.622 &  0.068   &  0.020           &   0.062   &    0.093   \\
 $  0.68<x_B<0.72$ &  2.092 &  0.080   &  0.028           &   0.044   &    0.096   \\
 $  0.72<x_B<0.76$ &  2.671 &  0.094   &  0.046           &   0.075   &    0.128   \\
 $  0.76<x_B<0.80$ &  3.102 &  0.104   &  0.071           &   0.140   &    0.189   \\
 $  0.80<x_B<0.84$ &  3.290 &  0.111   &  0.084           &   0.201   &    0.245   \\
 $  0.84<x_B<0.88$ &  2.953 &  0.106   &  0.065           &   0.144   &    0.190   \\
 $  0.88<x_B<0.92$ &  1.897 &  0.079   &  0.094           &   0.113   &    0.167   \\
 $  0.92<x_B<0.96$ &  0.753 &  0.042   &  0.051           &   0.205   &    0.215   \\
 $  0.96<x_B<1.00$ &  0.090 &  0.011   &  0.004           &   0.061   &    0.063   \\
\hline
\end{tabular}
\caption{
\label{table:final} 
The scaled $B$-hadron energy distribution.
}
\end{center}
\end{table}

\clearpage

The statistical correlation matrix is shown in Table~\ref{table:correl}.

\begin{table}[htb]
\footnotesize
\let\\=\cr
\def\hrulefill{\leaders\hrule\hfill}
\offinterlineskip
\hspace{-1cm}
\vbox{\halign{\vrule height 3ex depth 1.2ex\hfil# \vrule
& \hfil#
& \hfil#
& \hfil#
& \hfil#
& \hfil#
& \hfil#
& \hfil#
& \hfil#
& \hfil#
& \hfil#
& \hfil#
& \hfil#
& \hfil#
& \hfil#
& \hfil#
& \hfil#
& \hfil#
& \hfil#
& \hfil#
& \hfil#
& \hfil#
& \hfil# \vrule
\cr
\multispan{23}\hrulefill\\
{} Bin&4 &5 &6 &7 &8 &9 &10 &11 &12 &13 &14 &15 &16 &17 &18 &19 &20 &
21 &22 &23 &24 &25 \\%
\multispan{23}\hrulefill\\
5&69.4&&&&&&&&&&&&&&
&&&&&&&\\
6&42.0&79.0&&&&&&&&&&&&&
&&&&&&&\\
7&19.3&46.5&80.7&&&&&&&&&&&&
&&&&&&&\\
8&7.1&23.5&47.2&82.3&&&&&&&&&&&
&&&&&&&\\
9&6.9&10.6&18.3&39.6&79.0&&&&&&&&&&
&&&&&&&\\
10&4.8&6.0&9.8&20.9&48.5&82.1&&&&&&&&&
&&&&&&&\\
11&3.3&0.5&2.4&7.6&20.1&46.9&83.1&&&&&&&&
&&&&&&&\\
12&5.6&1.1&1.5&2.6&7.1&23.1&54.0&87.7&&&&&&&
&&&&&&&\\
13&5.8&0.1&-2.2&-2.5&-1.4&4.4&19.6&51.2&82.2&&&&&&
&&&&&&&\\
14&-0.3&-1.8&-2.8&-3.2&-2.9&-0.1&7.8&28.4&53.8&86.0&&&&&
&&&&&&&\\
15&-4.5&-4.6&-6.7&-6.8&-7.6&-7.5&-4.6&7.6&24.2&55.6&86.3&&&&
&&&&&&&\\
16&-7.2&-9.8&-12.6&-13.3&-13.7&-12.8&-13.3&-7.6&1.4&24.6&56.5&88.6&&&
&&&&&&&\\
17&-7.8&-11.5&-13.9&-15.4&-18.1&-19.2&-21.1&-19.1&-15.5&-3.1&18.4&55.4&85.8&&
&&&&&&&\\
18&-7.7&-15.0&-19.8&-22.6&-25.1&-26.1&-28.8&-29.1&-29.5&-24.8&-14.8&12.8&47.2&82.6&
&&&&&&&\\
19&-12.0&-19.5&-24.0&-28.1&-31.8&-33.4&-36.4&-38.0&-40.1&-39.5&-38.4&-24.4&1.3&39.1&80.6
&&&&&&&\\
20&-12.2&-22.9&-29.2&-33.9&-37.5&-38.8&-43.4&-48.2&-53.0&-55.3&-57.9&-50.8&-33.6&-5.3&37.5
&80.3&&&&&&\\
21&-15.3&-22.3&-27.3&-30.8&-33.8&-35.7&-40.3&-44.2&-47.4&-50.1&-57.7&-64.0&-62.5&-51.3&-19.8
&28.7&77.8&&&&&\\
 22&-13.9&-18.6&-22.7&-25.9&-28.8&-30.5&-34.5&-38.4&-42.0&-45.5&-52.0&-58.5&-60.7&-59.5&-45.2
&-12.1&39.2&85.6&&&&\\
23&-10.5&-13.1&-16.3&-18.8&-21.1&-22.3&-24.7&-27.8&-30.7&-33.9&-39.7&-46.3&-50.7&-54.6&-52.2
&-36.9&0.8&52.1&85.8&&&\\
24&-7.4&-9.3&-11.4&-13.0&-14.5&-15.5&-17.4&-19.8&-21.5&-23.8&-28.3&-33.5&-37.5&-41.7&-43.3
&-38.1&-15.8&23.9&57.7&89.1&&\\
25&-5.1&-6.4&-7.7&-8.1&-9.0&-10.3&-11.8&-13.7&-14.7&-16.0&-19.2&-22.9&-25.8&-28.7&-30.9
&-30.3&-19.0&6.2&32.9&68.8&91.2&\\
\multispan{23}\hrulefill\\
}}
\caption{
The statistical correlation matrix (\%).
}
\label{table:correl}
\end{table}

\clearpage

\section{Summary and Conclusions}

We have used the excellent tracking and vertexing capabilities of SLD 
to reconstruct the energies of $B$ hadrons in \ep \ra \z0 events over 
the full kinematic range by applying a new kinematic technique to 
an {\em inclusive} sample of reconstructed $B$-hadron
decay vertices.  The $B$ selection efficiency of the
method is 4.2\% and the resolution on the $B$ energy is about 9.6\% for 
roughly 83\% of the reconstructed decays.  The energy resolution for
low-energy $B$ hadrons is significantly better than in previous measurements.

We compared our measurement with several models of $b$-quark fragmentation.
The Bowler, Lund and Kartvelishivili \etal models, implemented within the
JETSET string fragmentation scheme, describe our data, as does the UCLA
model. None of the Braaten \etal, Collins-Spiller or Peterson \etal models implemented
within JETSET, nor HERWIG, describes the data.

The raw scaled $B$-energy distribution was corrected 
for bin-to-bin migrations caused by the resolution of the method, 
and for selection efficiency, to derive an estimate of the underlying
true distribution 
for weakly-decaying $B$ hadrons produced in \z0 decays.  
Systematic uncertainties in the correction were evaluated 
and found to be significantly smaller than those of previous 
direct $B$-energy measurements.  The final corrected $x_{B}$ distribution 
$D(x_{B})$ is shown in Figure~\ref{average}.
This result is consistent with, and supersedes, our previous 
measurements~\cite{sldbfrag,sldbfragprl}. It is also consistent with a recent precise
measurement~\cite{newaleph} 

It is conventional to evaluate the mean of this $B$-energy 
distribution, $<x_{B}>$.
For each of the seven parameter-dependent functions that provide a 
reasonable description of the data 
we evaluated $<x_{B}>$ from the distribution that corresponds to the 
optimised parameter(s); 
these are listed in Table~\ref{table:modelresult} and 
Table~\ref{table:formresult}.  For the UCLA model, which contains no
arbitrary parameters relating to $b$-quark fragmentation, 
we evaluated $<x_{B}>$ from the corresponding unfolded distribution 
shown in Fig.~\ref{overlay}; this yields 
$<x_{B}>$ = 0.712.
We took the average 
of the eight values of $<x_{B}>$ as our central value, and 
defined the model-dependent uncertainty to be the r.m.s. deviation.  
We obtained 
\begin{eqnarray}
<x_{B}>\quad=\quad 0.709\pm 0.003 (stat.)\pm 0.003 (syst.)\pm 0.002 (model),
\label{eqn:average}
\end{eqnarray}
It can be seen that $<x_{B}>$ is relatively insensitive to the variety of 
allowed forms of the shape of the fragmentation function.

\section*{Acknowledgements}
We thank the personnel of the SLAC accelerator department and the
technical
staffs of our collaborating institutions for their outstanding efforts
on our behalf.

\vskip .5truecm
\small
\vbox{\footnotesize\renewcommand{\baselinestretch}{1}\noindent
$^*$Work supported by Department of Energy
  contracts:

  DE-FG02-91ER40676 (BU),
  DE-FG03-91ER40618 (UCSB),
  DE-FG03-92ER40689 (UCSC),

  DE-FG03-93ER40788 (CSU),
  DE-FG02-91ER40672 (Colorado),
  DE-FG02-91ER40677 (Illinois),

  DE-AC03-76SF00098 (LBL),
  DE-FG02-92ER40715 (Massachusetts),
  DE-FC02-94ER40818 (MIT),

  DE-FG03-96ER40969 (Oregon),
  DE-AC03-76SF00515 (SLAC),
  DE-FG05-91ER40627 (Tennessee),

  DE-FG02-95ER40896 (Wisconsin),
  DE-FG02-92ER40704 (Yale);

  National Science Foundation grants:

  PHY-91-13428 (UCSC),
  PHY-89-21320 (Columbia),
  PHY-92-04239 (Cincinnati),

  PHY-95-10439 (Rutgers),
  PHY-88-19316 (Vanderbilt),
  PHY-92-03212 (Washington);

  The UK Particle Physics and Astronomy Research Council
  (Brunel, Oxford and RAL);

  The Istituto Nazionale di Fisica Nucleare of Italy

  (Bologna, Ferrara, Frascati, Pisa, Padova, Perugia);

  The Japan-US Cooperative Research Project on High Energy Physics
  (Nagoya, Tohoku);

  The Korea Research Foundation (Soongsil, 1997).}




\vfill
\eject

\section*{$^{**}$List of Authors}
%
%
%
\begin{center}
\def\iAOMORI{$^{(1)}$}
\def\iBRI{$^{(2)}$}
\def\iBRUN{$^{(3)}$}
\def\iBU{$^{(4)}$}
\def\iCOLO{$^{(5)}$}
\def\iCSU{$^{(6)}$}
\def\iFERR{$^{(7)}$}
\def\iFRAS{$^{(8)}$}
\def\iJHU{$^{(9)}$}
\def\iLBL{$^{(10)}$}
\def\iMASS{$^{(11)}$}
\def\iMISSI{$^{(12)}$}
\def\iMIT{$^{(13)}$}
\def\iMOSCOW{$^{(14)}$}
\def\iNAGO{$^{(15)}$}
\def\iOREG{$^{(16)}$}
\def\iOXF{$^{(17)}$}
\def\iPERU{$^{(18)}$}
\def\iRAL{$^{(19)}$}
\def\iRUTG{$^{(20)}$}
\def\iSLAC{$^{(21)}$}
\def\iSOONG{$^{(22)}$}
\def\iTENN{$^{(23)}$}
\def\iTOHO{$^{(24)}$}
\def\iUCSB{$^{(25)}$}
\def\iUCSC{$^{(26)}$}
\def\iVAND{$^{(27)}$}
\def\iWASH{$^{(28)}$}
\def\iWISC{$^{(29)}$}
\def\iYALE{$^{(30)}$}

  \baselineskip=.75\baselineskip
\mbox{Koya Abe\unskip,\iTOHO}
\mbox{Kenji Abe\unskip,\iNAGO}
\mbox{T. Abe\unskip,\iSLAC}
\mbox{I. Adam\unskip,\iSLAC}
\mbox{H. Akimoto\unskip,\iSLAC}
\mbox{D. Aston\unskip,\iSLAC}
\mbox{K.G. Baird\unskip,\iMASS}
\mbox{C. Baltay\unskip,\iYALE}
\mbox{H.R. Band\unskip,\iWISC}
\mbox{T.L. Barklow\unskip,\iSLAC}
\mbox{J.M. Bauer\unskip,\iMISSI}
\mbox{G. Bellodi\unskip,\iOXF}
\mbox{R. Berger\unskip,\iSLAC}
\mbox{G. Blaylock\unskip,\iMASS}
\mbox{J.R. Bogart\unskip,\iSLAC}
\mbox{G.R. Bower\unskip,\iSLAC}
\mbox{J.E. Brau\unskip,\iOREG}
\mbox{M. Breidenbach\unskip,\iSLAC}
\mbox{W.M. Bugg\unskip,\iTENN}
\mbox{D. Burke\unskip,\iSLAC}
\mbox{T.H. Burnett\unskip,\iWASH}
\mbox{P.N. Burrows\unskip,\iOXF}
\mbox{A. Calcaterra\unskip,\iFRAS}
\mbox{R. Cassell\unskip,\iSLAC}
\mbox{A. Chou\unskip,\iSLAC}
\mbox{H.O. Cohn\unskip,\iTENN}
\mbox{J.A. Coller\unskip,\iBU}
\mbox{M.R. Convery\unskip,\iSLAC}
\mbox{V. Cook\unskip,\iWASH}
\mbox{R.F. Cowan\unskip,\iMIT}
\mbox{G. Crawford\unskip,\iSLAC}
\mbox{C.J.S. Damerell\unskip,\iRAL}
\mbox{M. Daoudi\unskip,\iSLAC}
\mbox{N. de Groot\unskip,\iBRI}
\mbox{R. de Sangro\unskip,\iFRAS}
\mbox{D.N. Dong\unskip,\iMIT}
\mbox{M. Doser\unskip,\iSLAC}
\mbox{R. Dubois\unskip,}
\mbox{I. Erofeeva\unskip,\iMOSCOW}
\mbox{V. Eschenburg\unskip,\iMISSI}
\mbox{E. Etzion\unskip,\iWISC}
\mbox{S. Fahey\unskip,\iCOLO}
\mbox{D. Falciai\unskip,\iFRAS}
\mbox{J.P. Fernandez\unskip,\iUCSC}
\mbox{K. Flood\unskip,\iMASS}
\mbox{R. Frey\unskip,\iOREG}
\mbox{E.L. Hart\unskip,\iTENN}
\mbox{K. Hasuko\unskip,\iTOHO}
\mbox{S.S. Hertzbach\unskip,\iMASS}
\mbox{M.E. Huffer\unskip,\iSLAC}
\mbox{X. Huynh\unskip,\iSLAC}
\mbox{M. Iwasaki\unskip,\iOREG}
\mbox{D.J. Jackson\unskip,\iRAL}
\mbox{P. Jacques\unskip,\iRUTG}
\mbox{J.A. Jaros\unskip,\iSLAC}
\mbox{Z.Y. Jiang\unskip,\iSLAC}
\mbox{A.S. Johnson\unskip,\iSLAC}
\mbox{J.R. Johnson\unskip,\iWISC}
\mbox{R. Kajikawa\unskip,\iNAGO}
\mbox{M. Kalelkar\unskip,\iRUTG}
\mbox{H.J. Kang\unskip,\iRUTG}
\mbox{R.R. Kofler\unskip,\iMASS}
\mbox{R.S. Kroeger\unskip,\iMISSI}
\mbox{M. Langston\unskip,\iOREG}
\mbox{D.W.G. Leith\unskip,\iSLAC}
\mbox{V. Lia\unskip,\iMIT}
\mbox{C. Lin\unskip,\iMASS}
\mbox{G. Mancinelli\unskip,\iRUTG}
\mbox{S. Manly\unskip,\iYALE}
\mbox{G. Mantovani\unskip,\iPERU}
\mbox{T.W. Markiewicz\unskip,\iSLAC}
\mbox{T. Maruyama\unskip,\iSLAC}
\mbox{A.K. McKemey\unskip,\iBRUN}
\mbox{R. Messner\unskip,\iSLAC}
\mbox{K.C. Moffeit\unskip,\iSLAC}
\mbox{T.B. Moore\unskip,\iYALE}
\mbox{M. Morii\unskip,\iSLAC}
\mbox{D. Muller\unskip,\iSLAC}
\mbox{V. Murzin\unskip,\iMOSCOW}
\mbox{S. Narita\unskip,\iTOHO}
\mbox{U. Nauenberg\unskip,\iCOLO}
\mbox{H. Neal\unskip,\iYALE}
\mbox{G. Nesom\unskip,\iOXF}
\mbox{N. Oishi\unskip,\iNAGO}
\mbox{D. Onoprienko\unskip,\iTENN}
\mbox{L.S. Osborne\unskip,\iMIT}
\mbox{R.S. Panvini\unskip,\iVAND}
\mbox{C.H. Park\unskip,\iSOONG}
\mbox{I. Peruzzi\unskip,\iFRAS}
\mbox{M. Piccolo\unskip,\iFRAS}
\mbox{L. Piemontese\unskip,\iFERR}
\mbox{R.J. Plano\unskip,\iRUTG}
\mbox{R. Prepost\unskip,\iWISC}
\mbox{C.Y. Prescott\unskip,\iSLAC}
\mbox{B.N. Ratcliff\unskip,\iSLAC}
\mbox{J. Reidy\unskip,\iMISSI}
\mbox{P.L. Reinertsen\unskip,\iUCSC}
\mbox{L.S. Rochester\unskip,\iSLAC}
\mbox{P.C. Rowson\unskip,\iSLAC}
\mbox{J.J. Russell\unskip,\iSLAC}
\mbox{O.H. Saxton\unskip,\iSLAC}
\mbox{T. Schalk\unskip,\iUCSC}
\mbox{B.A. Schumm\unskip,\iUCSC}
\mbox{J. Schwiening\unskip,\iSLAC}
\mbox{V.V. Serbo\unskip,\iSLAC}
\mbox{G. Shapiro\unskip,\iLBL}
\mbox{N.B. Sinev\unskip,\iOREG}
\mbox{J.A. Snyder\unskip,\iYALE}
\mbox{H. Staengle\unskip,\iCSU}
\mbox{A. Stahl\unskip,\iSLAC}
\mbox{P. Stamer\unskip,\iRUTG}
\mbox{H. Steiner\unskip,\iLBL}
\mbox{D. Su\unskip,\iSLAC}
\mbox{F. Suekane\unskip,\iTOHO}
\mbox{A. Sugiyama\unskip,\iNAGO}
\mbox{A. Suzuki\unskip,\iNAGO}
\mbox{M. Swartz\unskip,\iJHU}
\mbox{F.E. Taylor\unskip,\iMIT}
\mbox{J. Thom\unskip,\iSLAC}
\mbox{E. Torrence\unskip,\iMIT}
\mbox{T. Usher\unskip,\iSLAC}
\mbox{J. Va'vra\unskip,\iSLAC}
\mbox{R. Verdier\unskip,\iMIT}
\mbox{D.L. Wagner\unskip,\iCOLO}
\mbox{A.P. Waite\unskip,\iSLAC}
\mbox{S. Walston\unskip,\iOREG}
\mbox{A.W. Weidemann\unskip,\iTENN}
\mbox{E.R. Weiss\unskip,\iWASH}
\mbox{J.S. Whitaker\unskip,\iBU}
\mbox{S.H. Williams\unskip,\iSLAC}
\mbox{S. Willocq\unskip,\iMASS}
\mbox{R.J. Wilson\unskip,\iCSU}
\mbox{W.J. Wisniewski\unskip,\iSLAC}
\mbox{J.L. Wittlin\unskip,\iMASS}
\mbox{M. Woods\unskip,\iSLAC}
\mbox{T.R. Wright\unskip,\iWISC}
\mbox{R.K. Yamamoto\unskip,\iMIT}
\mbox{J. Yashima\unskip,\iTOHO}
\mbox{S.J. Yellin\unskip,\iUCSB}
\mbox{C.C. Young\unskip,\iSLAC}
\mbox{H. Yuta\unskip.\iAOMORI}

\it
  \vskip \baselineskip                   
  \baselineskip=.75\baselineskip   
\iAOMORI
  Aomori University, Aomori, 030 Japan, \break
\iBRI
  University of Bristol, Bristol, United Kingdom, \break
\iBRUN
  Brunel University, Uxbridge, Middlesex, UB8 3PH United Kingdom, \break
\iBU
  Boston University, Boston, Massachusetts 02215, \break
\iCOLO
  University of Colorado, Boulder, Colorado 80309, \break
\iCSU
  Colorado State University, Ft. Collins, Colorado 80523, \break
\iFERR
  INFN Sezione di Ferrara and Universita di Ferrara, I-44100 Ferrara, Italy,
\break
\iFRAS
  INFN Laboratori Nazionali di Frascati, I-00044 Frascati, Italy, \break
\iJHU
  Johns Hopkins University,  Baltimore, Maryland 21218-2686, \break
\iLBL
  Lawrence Berkeley Laboratory, University of California, Berkeley, California
94720, \break
\iMASS
  University of Massachusetts, Amherst, Massachusetts 01003, \break
\iMISSI
  University of Mississippi, University, Mississippi 38677, \break
\iMIT
  Massachusetts Institute of Technology, Cambridge, Massachusetts 02139, \break
\iMOSCOW
  Institute of Nuclear Physics, Moscow State University, 119899 Moscow, Russia,
\break
\iNAGO
  Nagoya University, Chikusa-ku, Nagoya, 464 Japan, \break
\iOREG
  University of Oregon, Eugene, Oregon 97403, \break
\iOXF
  Oxford University, Oxford, OX1 3RH, United Kingdom, \break
\iPERU
  INFN Sezione di Perugia and Universita di Perugia, I-06100 Perugia, Italy,
\break
\iRAL
  Rutherford Appleton Laboratory, Chilton, Didcot, Oxon OX11 0QX United Kingdom,
\break
\iRUTG
  Rutgers University, Piscataway, New Jersey 08855, \break
\iSLAC
  Stanford Linear Accelerator Center, Stanford University, Stanford, California
94309, \break
\iSOONG
  Soongsil University, Seoul, Korea 156-743, \break
\iTENN
  University of Tennessee, Knoxville, Tennessee 37996, \break
\iTOHO
  Tohoku University, Sendai, 980 Japan, \break
\iUCSB
  University of California at Santa Barbara, Santa Barbara, California 93106,
\break
\iUCSC
  University of California at Santa Cruz, Santa Cruz, California 95064, \break
\iVAND
  Vanderbilt University, Nashville,Tennessee 37235, \break
\iWASH
  University of Washington, Seattle, Washington 98105, \break
\iWISC
  University of Wisconsin, Madison,Wisconsin 53706, \break
\iYALE
  Yale University, New Haven, Connecticut 06511. \break

\rm
%

\end{center}

\vskip 1truecm
 
\clearpage

\clearpage

\clearpage

\clearpage

\clearpage

\clearpage

\clearpage

\clearpage

\clearpage

\clearpage

\end{document}